\documentclass[12pt]{iopart}
\usepackage[dvips]{graphics}
\usepackage[dvips]{color}
  
\begin{document}

\title[Resonant X-ray Scattering in Manganites]
{Resonant X-ray Scattering in Manganites \\
- Study of Orbital Degree of Freedom -}

\author{Sumio Ishihara and Sadamichi Maekawa}

\address{\dag\ Department of Applied Physics, University of Tokyo, Tokyo 113-8656, Japan}

\address{\ddag\ Institute for Materials Research, Tohoku University, Sendai 980-8577, Japan}

\address{E-mail: ishihara@ap.t.u-tokyo.ac.jp and maekawa@imr.tohoku.ac.jp}

\begin{abstract}
Orbital degree of freedom of electrons 
and its interplay  with spin, charge and lattice degrees of freedom 
are one of the central issues in colossal magnetoresistive manganites. 
The orbital degree of freedom has until recently remained hidden, since 
it does not couple directly to most of experimental probes. 
Development of synchrotron light sources has changed the 
situation; by the resonant x-ray scattering (RXS) technique 
the orbital ordering has successfully been observed . 
In this article, we review progress in the recent studies of RXS 
in manganites. We start with a detailed review of the RXS experiments 
applied to the orbital ordered manganites and other correlated electron systems. 
We derive the scattering cross section of RXS where the tensor character of the 
atomic scattering factor (ASF) with respect to the x-ray polarization is stressed. 
Microscopic mechanisms of the anisotropic tensor character of ASF 
is introduced and numerical results of ASF and the 
scattering intensity are presented. 
The azimuthal angle scan is a unique experimental method to 
identify RXS from the orbital degree of freedom. 
A theory of the azimuthal angle and polarization 
dependence of the RXS intensity is presented.  
The theoretical results show good agreement with the experiments in manganites. 
Apart from the microscopic description of ASF, 
a theoretical framework of RXS to relate directly to 
the $3d$ orbital is presented. 
The scattering cross section is represented by 
the correlation function of the pseudo-spin operator for the orbital degree of freedom. 
A theory is extended to the resonant inelastic x-ray scattering 
and methods to observe excitations of the orbital degree of freedom are proposed. 
\end{abstract}


\submitto{\RPP}
\maketitle

\tableofcontents
\clearpage

\section{Introduction}
\label{sec:intro}
\begin{figure}[t]
\begin{center}
        \resizebox{0.5\linewidth}{!}{\includegraphics{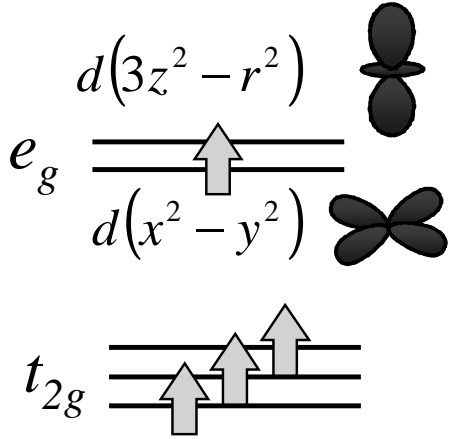}}
\end{center}
        \caption{A schematic picture of the energy level in a Mn$^{3+}$ ion 
        and the $3d_{3z^2-r^2}$ and $3d_{x^2-y^2}$ orbitals.}
        \label{fig:eg}
\end{figure}
The discovery of the high Tc superconducting cuprates \cite{bednortz} have stimulated 
the revival studies of a variety of transition-metal oxides 
from modern theoretical and experimental view points. 
Among them, special attention is placed on manganites $A_{1-x}$$B_x$MnO$_3$ \cite{jonker,jonker2}
with pseudo-cubic perovskite structure  
and their related materials. 
Here, $A$ and $B$ are trivalent and divalent cations, respectively. 
This is owing to the newly discovered colossal magnetoresistance (CMR)\cite{chahara,helmolt,tokura95,jin}: 
a huge decreasing of the electrical resistivity by applying a magnetic field. 
As well as CMR, 
a number of dramatic and fruitful phenomena in magnetic, electric and optical properties 
has been reported. 
Some of them can not be explained by the simple double exchange scenario \cite{millis1,millis2,roder,inoue} and 
are ascribed to the strong interplay between 
spin, charge and orbital degrees of freedom of an electron, as well as lattice 
\cite{review,review_2,review_3,review_5,review_4}. 
Let us compare the electronic structure in CMR manganites with 
that in high Tc cuprates. 
As well as cuprates, 
manganites are known to be strongly correlated electron systems. 
In particular, the strong Hund coupling is stressed 
for origin of the ferromagnetic metallic state. 
This is so-called the double exchange interaction \cite{zener,anderson,degenne,kubo1,kubo2,furukawa1,furukawa2}.  
The electron configuration of a Cu$^{2+}$ ion in La$_2$CuO$_4$ is $3d^{9}$ 
and it is located at a center of a tetragonally distorted CuO$_6$ octahedron. 
Here, one hole occupies the $3d_{x^2-y^2}$ orbital. 
On the other hand, 
four $3d$ electrons in a Mn$^{3+}$ ion form the high spin state,   
and the electrons occupy the three $t_{2g}$ orbitals and 
one of the doubly degenerate $e_g$ orbitals as $t_{2g}^{3}e_g^{1}$ (see Fig.~\ref{fig:eg}). 
Therefore, this ion has a degree of freedom indicating 
which $e_g$ orbital is occupied by an electron. 
This is termed the orbital degree of freedom 
and is recognized as the third degree of freedom of an electron, 
in addition to the spin and charge ones. 

\begin{figure}[t]
\begin{center}
        \resizebox{0.6\linewidth}{!}{\includegraphics{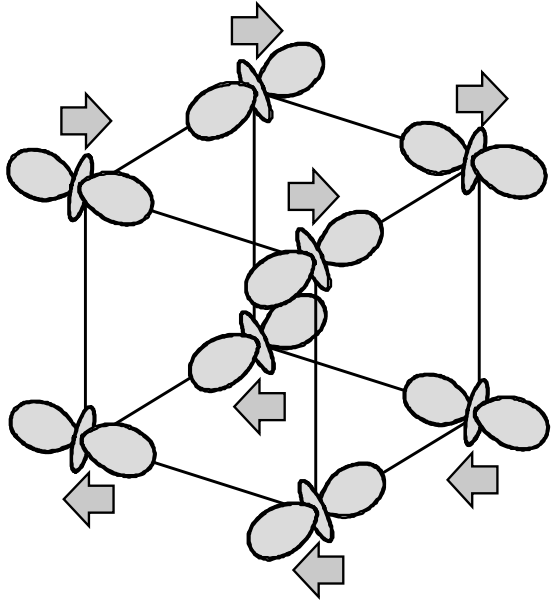}}
\end{center}
        \caption{The orbital ordered state in LaMnO$_3$.
        Bold arrows indicate Mn spins.}
        \label{fig:orbilmo3}
\end{figure}
In solid, the orbital degree of freedom often shows the long range ordering, 
termed the orbital ordering, 
which is caused by the inter-site interaction between orbitals. 
One of the examples is found in the parent compound of CMR manganites, LaMnO$_3$,  
where $3d_{3x^2-r^2}$ and $3d_{3y^2-r^2}$ orbitals are alternately aligned 
in the $ab$ plane below 780K (see Fig.~\ref{fig:orbilmo3}) 
\cite{elemans,matsumoto_70,goodenough,mitchell96,huang,rodriguez}. 
It is widely accepted that 
this orbital ordering governs the anisotropic spin order 
termed the $A$-type antiferromagnetic (AF) order 
where spins are ferromagnetically aligned in the $ab$ plane and 
antiferromagnetically along the $c$ axis \cite{goodenough,wollan,kugel73,hamada,mizokawa,
ishihara_eh,ishihara_eh2,millis_JT,solovyev}. 
Another type of the orbital ordering appears in $A_{0.5}B_{0.5}$MnO$_3$ 
where a nominal valence of a Mn ion is 3.5+ \cite{wollan,jirak80,tomioka,hwang,mori,koshibae,anisimov,solovyev_co}. 
Below a certain temperature, 
Mn$^{3+}$ and Mn$^{4+}$ ions spatially order 
and, in the Mn$^{3+}$ sublattice, 
two kinds of orbital sublattice appear. 
The charge and orbital orderings occur at the same temperature 
and, at finally, the spin order termed the CE-type AF order is realized. 
Thus, all electronic degrees of freedom are frozen in the low temperature phase. 
Since CMR appears at the vicinity of the charge and orbital ordering transition, 
it is recognized that the orbital plays a crucial role on  
CMR and other dramatic phenomena in manganites. 

Theoretical study of the electronic orbital in 3$d$ electron systems 
is retrospective to more than 40 years ago. 
Signs and magnitudes of the superexchange interactions  
have been studied in the systems with orbital degeneracy in 1950's 
\cite{vanvleck,goodenough,kanamori59}. 
The orbital degeneracy brings about the ferromagnetic interaction 
between same kinds of ions 
in the case where an angle of the bond connecting the nearest neighboring (NN) 
magnetic ions is 180$^\circ$. 
As well as the superexchange interaction between spins, 
the virtual exchanges of electrons under 
the strong Coulomb interactions induce the inter-site interactions between orbitals. 
Being based on the theoretical models, where spin and orbital are treated on an equal footing, 
the mutual relations between spin and orbital orderings were  
investigated in several magnetic compounds 
\cite{roth,cyrot,kugel72,kugel76,khomskii73,korovin,komarov,inagaki}.
On the other hand, 
the orbital ordering has been studied in connection with Jahn-Teller effects. 
In molecules, a degeneracy of electronic orbitals is always lifted spontaneously by lowering 
symmetry of molecule except for chain-shaped molecules.  
This is the Jahn-Teller theorem \cite{jt}. 
In solid, the virtual exchange of phonons brings about the inter-site interaction between orbitals and 
causes the orbital ordering associated with the structural phase transition. 
This is termed the cooperative Jahn-Teller effects \cite{kanamori60,kataoka72,jtreview,jtreview2}. 
After the discovery of CMR, 
theoretical study of the orbital degree of freedom has been revived and, in particular, 
the following new points of researches are focused on: 
(1) dynamics of the orbital degree of freedom 
\cite{ishihara_eh,ishihara_eh2,feiner0,khaliullin,feiner,khaliullin2,brink,allen,perebeinos}, 
(2) roles of the orbital on magnetic, transport and optical properties in metals 
\cite{ishihara_liq,shiba,kilian,mack,nakano,maezono_1,maezono_2,veenendaal,sheng,okamoto_pt,koshibae2,
allen_pe,brink_pe,bala_pe,yin}, 
and 
(3) exotic orbital ordered states 
\cite{kilian_99,yunoki,okamoto_ps,hotta,khomskii_cm,takahashi_cm,maezono_cm}.  
Actually, a variety of orbital states was proposed theoretically 
in manganites and other transition-metal oxides. 
However, direct comparisons of these predictions with experimental results have 
been limited up to recently, 
because the orbital degree of freedom does not couple directly to most of experimental probes. 
This is in contrast to the spin and charge degrees of freedom 
which are directly detected by the neutron and electron/x-ray diffraction 
techniques, respectively.  
 
In 1998, Murakami and coworkers applied the resonant x-ray scattering (RXS) method to 
one of the orbital ordered manganites La$_{0.5}$Sr$_{1.5}$MnO$_4$ 
and successfully observed the orbital order \cite{murakami_214}. 
This technique has been rapidly developed and expanded to several compounds with orbital degree of freedom.  
Nowadays, RXS uncovers a whole feature of the spin-charge-orbital coupled systems. 
In this article,  
we review the recent progress of the theoretical and experimental developments of RXS in manganites. 

In Sec.~\ref{sec:histry}, histories of the observation of the orbital and RXS are 
introduced. The recent experimental results of RXS are reviewed in Sec.~\ref{sec:recentex}. 
Section~\ref{sec:theory} is devoted to the review of the theoretical study of RXS. 
We derive the scattering cross section of RXS and stress the anisotropy of the 
atomic scattering factor (ASF) in Sec.~\ref{sec:scs}. 
Microscopic mechanisms of the anisotropy of 
ASF in orbital ordered state is introduced  
in Sec.~\ref{sec:mechanism}. 
The azimuthal angle scan is a unique experimental method by which RXS from the orbital ordering is identified. 
A theory of the azimuthal angle and polarization dependence of RXS is presented in Sec.~\ref{sec:azimuth}. 
Apart from the micropcopic mechanisms of the anisotropy of ASF, 
we introduce, in Sec.~\ref{sec:group}, 
the general theoretical framework of RXS where the scattering cross section is 
directly related to the orbital degree of freedom. 
This is applied to study 
of the orbital ordering, fluctuation and excitations. 
Section~\ref{sec:summary} is devoted to the summary. 

\section{Survey of experiments}
\label{sec:exp}
\subsection{Histrical background of the observation of orbital and RXS}
\label{sec:histry}

Experimental observation of the orbital degree of freedom of an electron is, in general, a hard task, 
unlike spin, charge and lattice degrees of freedom. 
As mentioned previously, in molecules, a degeneracy of electronic orbitals is lifted by the Jahn-Teller effects. 
It may be expected that, also in solid, 
the orbital ordering always occurs by a crystal lattice distortion
as known to be the cooperative Jahn-Teller effects. 
Even when the orbital ordering is caused by other mechanisms, 
the crystal lattice may follow a shape of the electronic cloud by the electron-lattice interactions.  
Thus, one may think that 
the orbital ordering can be observed through measuring the lattice distortion. 
However, this is not always true and  
one to one correspondence between orbital and lattice  
does not exist in solid. 
This is because there is an infinite number of degrees of freedom 
and several interactions acting on orbital and lattice 
compete and cooperate with each other.  

In principle, the scattering intensities in the several diffraction experiments 
depend on 
the anisotropic shape of the electronic wave function and/or electronic charge distribution. 
However, since the orbital dependence of the scattering intensity is 
too small to be observed easily, 
precise experiments in a wide range of diffraction angles 
and detailed data analyses are required. 
As an example, let us consider the conventional x-ray diffraction experiments. 
The peak intensity is proportional to the square of the electronic charge valence. 
On the other hand, a difference of the electronic orbital corresponds to a difference of the charge valence 
being less than a charge unit $e$. 
Thus, the orbital effects in the x-ray scattering intensity 
is usually smeared out by the scatterings from heavy ions, impurities, defects and so on. 
The first direct observation of the orbital ordered state was 
carried out by the polarized neutron diffraction method in K$_2$CuF$_4$ \cite{itoh,akimitsu}. 
This technique is applicable to the observation of orbital ordering in magnetically ordered state 
where both the spin and orbital moments are ascribed to the same $d$ or $f$ electrons. 
The magnetic structure factors $f_M$ which include information of the electronic orbital 
are measured at several reflection points and 
are compared with the model calculations. 
This method was recently applied to the ferromagnetic YTiO$_3$ and 
revealed successfully the type of the orbital ordered state \cite{ichikawa}. 
Another experimental observation of the orbital ordering 
was carried out by the charge density distribution study 
using the x-ray and/or electron diffraction technique. 
The charge density of a Cu$^{+}$ ion in Cu$_2$O 
was recently observed by utilizing the convergent-beam electron diffraction 
combined with x-ray diffraction \cite{zuo}. 
In general, x-ray scattering is sensitive to defects contained in a crystal and 
the scattering from defects is greater than the orbital dependent scattering. 
Thus, a perfect region of a crystal is selected by the electron microscopy 
and the charge distribution map is obtained by 
the electron and x-ray diffraction data in a wide region of scattering angles. 
An electronic charge-distribution is also obtained by 
the combined method of the Rietveld refinement of the x-ray powder diffraction data 
and the maximum entropy method (MEM). 
Here, MEM is utilized to construct a real space charge-distribution  
from the x-ray diffraction data by the Fourier transformation. 
This method has been applied to the structural studies of several fullerene and silicides, 
\cite{bricogne,takata0} 
and was recently applied to NdSr$_{2}$Mn$_2$O$_7$ \cite{takata}. 
\begin{figure}[t]
\begin{center}
        \resizebox{0.6\linewidth}{!}{\includegraphics{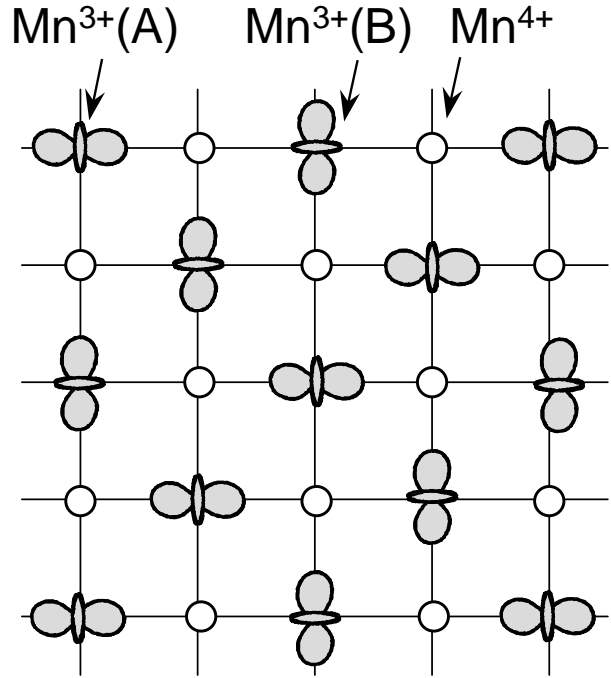}}
\end{center}
        \caption{The charge and orbital ordered 
        state in the $ab$ plane of 
        La$_{1.5}$Sr$_{0.5}$MnO$_4$. Open circles indicate Mn$^{4+}$ ions \cite{murakami_214}.}
\label{fig:coorder}
\end{figure}
\par
The study of the resonant x-ray scattering and 
anisotropic character of the x-ray scattering have been developed in the research field of 
the crystarography more than twenty years ago. 
The anisotropies of absorption and reflection in the visible 
light regions, i.e. the linear dichroism and birefringence (double reflection),
are well known phenomena and useful to the study of the electronic structure in solid. 
However, in the x-ray region, these are usually small and the scattering factor is 
treated as a scalar with respect to the polarization of x ray. 
The observations of the polarization dependent x-ray absorption and 
double reflection have been done by Templeton and Templeton 
near the V $K$ edge in crystalline VO(C$_5$H$_7$O$_2$)$_2$ \cite{templeton1}. 
The anisotropic feature was also observed in the diffraction experiments 
in Sodium Uranyl acetate, NaUO$_2$(O$_2$CCH$_3$)$_3$, 
and was found to be dramatically enhanced when the incident x-ray energy is tuned around U $L$ edge \cite{templeton2}. 
They have measured the polarization dependence of the diffraction intensity 
as a function of the x-ray energy and derived the tensor elements of the x-ray scattering factors. 
The crystal structure of this compound is cubic and does not 
exhibit dichroism and birefringence on a macroscopic scale. 
Therefore, this is caused by a lack of the cubic symmetry around U ion which 
is located on threefold axes. 
With these works as a start, 
a number of experimental and theoretical studies about x-ray dichroism and birefringence have been performed 
\cite{templeton3,templeton4,kolpakov,dmitrienko1,dmitrienko2,
petcov,kirfel,tsuji,nagano,reviewrxs}. 
In particular, Dmitrienko developed this issue theoretically 
from the view point of the forbidden reflection \cite{dmitrienko1,dmitrienko2}. 
He expressed the x-ray scattering factor by the anisotropic x-ray susceptibility tensor 
and derived new extinction rules being valid near the absorption edge. 
This reflection is termed the anisotropy of tensor of susceptibility (ATS) reflection. 
In the microscopic point of view, these phenomena are attributed to 
the chemical bonding of the edge atom in an anisotropic chemical environment. 
This brings about the orientation of unoccupied electronic levels. 

\begin{figure}[t]
\begin{center}
        \resizebox{0.7\linewidth}{!}{\includegraphics{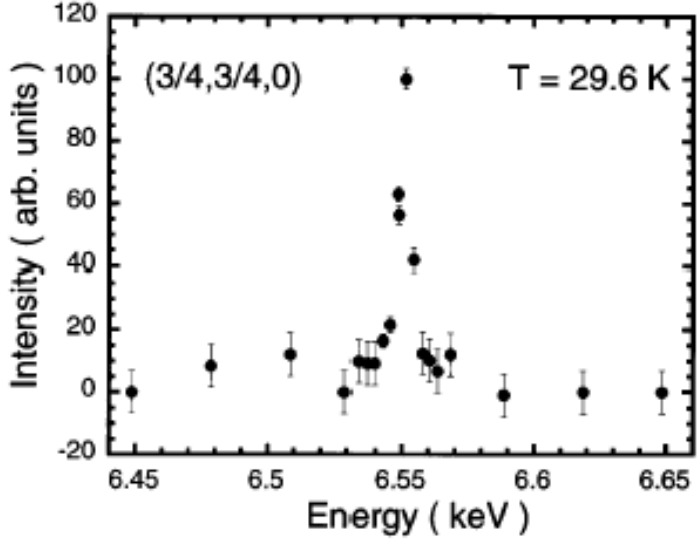}}
\end{center}
        \caption{Energy dependence of the scattering intensity in  
        La$_{1.5}$Sr$_{0.5}$MnO$_4$ 
        at the orbital superlattice reflection (3/4 3/4 0) \cite{murakami_214}.}
        \label{fig:214int}
\end{figure}
\subsection{A variety of recent RXS experiments}
\label{sec:recentex}
In 1998, the resonant x-ray scattering was first applied to study of the orbital ordering in 
one of the manganites La$_{0.5}$Sr$_{1.5}$MnO$_4$ \cite{murakami_214}. 
The formal valence of a Mn ion in this compound is 3.5+, that is, 
an average number of the $e_g$ electron is 0.5 per ion. 
Before the RXS experiment,  
an alternating charge ordering  
with a $\sqrt{2}a \times \sqrt{2} a \times c$ 
unit cell was suggested by the electron diffraction experiments below 270K \cite{moritomo,bao}. 
In addition, the magnetically ordered structure with a 
$2\sqrt{2}a \times 2\sqrt{2} a \times 2c$ unit cell was observed below 110K \cite{sternlieb}. 
This is the so-called $CE$-type AF ordering associated with 
the charge ordering and has been already observed in manganites with pseudo-cubic 
perovskite structure. 
More than forty years ago, in order to explain the $CE$-type AF  
ordering observed in La$_{0.5}$Ca$_{0.5}$MnO$_3$, an alternate ordering of 
the chemical bonds between Mn and O ions, termed the covalent-bond ordering, was proposed \cite{goodenough,wollan}. 
Through the systematic structural and magnetic studies by x-ray and neutron scatterings, 
this idea was reinforced, and the ordering of $e_g$ orbital associated 
with distorted MnO$_6$ octahedra was supposed \cite{jirak80}. 
Therefore,  
it was expected in La$_{0.5}$Sr$_{1.5}$MnO$_4$ that the similar multi-component ordering is realized, 
although there was no direct evidence of the orbital ordering.  

\begin{figure}[t]
\begin{center}
        \resizebox{0.6\linewidth}{!}{\includegraphics{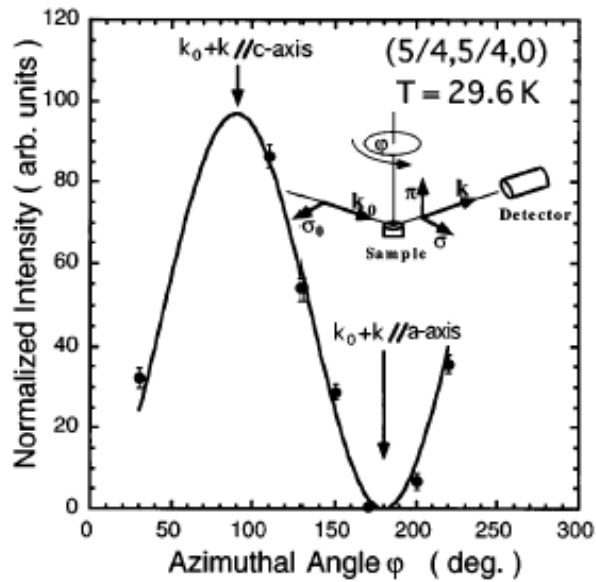}}
\end{center}
        \caption{Azimuthal angle dependence of the scattering intensity in La$_{1.5}$Sr$_{0.5}$MnO$_4$ 
        at the orbital superlattice reflection (5/4 5/4 0) \cite{murakami_214}.}
        \label{fig:214az}
\end{figure}
Murakami and coworkers applied RXS to investigate charge and orbital orderings in 
La$_{0.5}$Sr$_{1.5}$MnO$_4$ \cite{murakami_214}. 
They utilized the following characteristics in this compound and RXS technique: 
(1) The expected orbital ordering has a unit cell with  
$\sqrt{2}a \times 2\sqrt{2} a \times c$ which is distinct from the unit cells for the spin and charge orderings. 
A schematic picture of the charge and orbital ordered state is shown in Fig.~\ref{fig:coorder}. 
(2) ASF in RXS is a tensor with respect to polarization of x ray. 
This is available to detect the asphericity of the electron density. 
(3) A magnitude of ASF in RXS is sensitive to a valence of ion. 
To detect the orbital ordering, 
the incident x-ray energy was tuned near the Mn $K$ edge of 6.552KeV and 
the reflection at $({2l+1 \over 4}\ {2l+1 \over 4}\ n)$ was chosen. 
This reflection corresponds to a unit cell of the orbital ordering and termed 
orbital superlattice reflection. 
The structure factor at this point 
is represented by 
\begin{equation}
F\bigl({\textstyle \frac{2l+1}{4}} \ {\textstyle \frac{2l+1}{4}}\ 0 \bigr)=f({\rm Mn^{3+}(A)})-f({\rm Mn^{3+}(B)}) ,
\end{equation}
where $f({\rm Mn^{3+}(A)})$ and $f({\rm Mn^{3+}(B)})$ are ASF's in Mn$^{3+}$ with orbitals A and B, 
respectively. 
When all Mn$^{3+}$ ions are equivalent,  
this reflection disappears due to the structure factor. 
This is the extinction rule in the x ray scattering. 
The experimental results of the incident x-ray energy dependence of the scattering intensity 
are presented in Fig.~\ref{fig:214int}. 
The scattering intensity resonantly appears near the $K$-absorption edge, that is, 
the extinction rule is broken near the edge. 
This forbidden reflection is not ascribed to the normal part of ASF, 
unlike the well known case of the $(2l+1\ 2m+1\ 2n+1)$ and $(2l\ 2m\ 2n)$ reflections 
in Ge due to the anharmonic lattice vibration, 
because of the following two reasons; 
(1) The scattering intensity shows a resonant behavior 
near the absorption edge.  
(2) The normal part of ASF is given by  
\begin{equation}
f_{0A(B)}(\vec K) \propto \int e^{i \vec K \cdot \vec r} \rho_{A(B)}(\vec r) , 
\label{eq:rhoxy}
\end{equation}
where $f_{0A(B)}(\vec K)$ is the normal part of ASF in a Mn$^{3+}$ ion with A(B) orbital 
and $\rho_{A(B)}(\vec r)$ is its charge distribution. 
Because of the symmetry of the charge distribution  as  
\begin{equation}
\rho_A(x,y,z)=\rho_B(y,x,z) , 
\end{equation}
a difference of the normal part $f_{0A}(\vec K)-f_{0B}(\vec K)$ vanishes at $\vec K=({2l+1 \over 4}\ {2l+1 \over 4}\ 0)$. 

\begin{figure}[t]
\begin{center}
        \resizebox{0.7\linewidth}{!}{\includegraphics{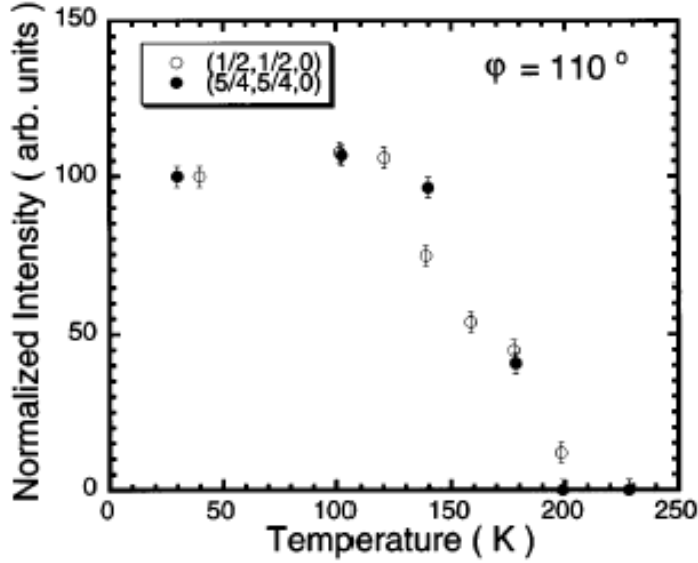}}
\end{center}
        \caption{Temperature dependence of the scattering 
        intensity in La$_{1.5}$Sr$_{0.5}$MnO$_4$ 
        at the orbital superlattice reflection (5/4 5/4 0) and that at the charge superlattice 
        reflection (1/2 1/2 0) \cite{murakami_214}.}
        \label{fig:214temp}
\end{figure}
The measurement of the angular dependence around the scattering vector, termed the azimuthal angle 
dependence, was utilized to confirm that this scattering results from the anomalous part of ASF. 
The x-ray polarization coming from the synchrotron light source is highly polarized, 
such as the $\sigma$ polarization in this case. 
Thus, the relative direction of electric vector to a sample is changed by rotating a sample around the scattering vector. 
The scattering due to the normal part of ASF does not show this dependence, since the ASF is scalar. 
The experimental results are shown in Fig.~\ref{fig:214az}. 
The scattering intensity oscillates as a function of the azimuthal angle $\varphi$; 
the intensity becomes its maximum (minimum) at $\varphi=90^\circ$ (180$^\circ$) 
where the incident x-ray polarization is parallel to the $ab$ plane ($c$ axis) of the sample. 
This dependence was fitted by a function $\sin^2 \varphi$ derived 
from the phenomenological calculation assuming the orbital ordering. 
Therefore, the forbidden reflections ascribed to the multiple scattering, such as the Ranninger scattering, is ruled out. 
\begin{figure}[t]
\begin{center}
        \resizebox{0.6\linewidth}{!}{\includegraphics{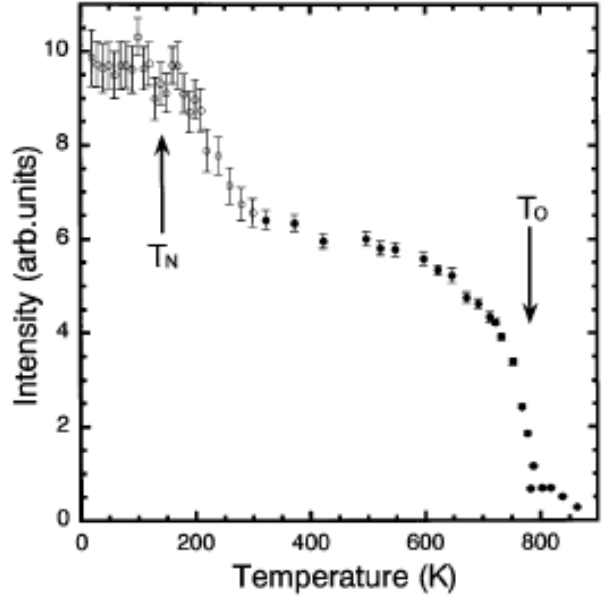}}
\end{center}
        \caption{Temperature dependence of the scattering intensity in LaMnO$_3$ 
        at the orbital superlattice reflection (3 0 0) \cite{murakami_113}.}
        \label{fig:orbilmo3t}
\end{figure}
It was concluded that the observed reflection originates from the anisotropic 
character of the anomalous part of ASF. 
The authors also measured the reflection at $({2l+1 \over 2}\ {2l+1 \over 2}\ 0)$
being proportional to the difference between ASF's of Mn$^{3+}$ and Mn$^{4+}$: 
\begin{equation}
F\bigl({\textstyle \frac{2l+1}{2}}\ {\textstyle \frac{2l+1}{2}}\ 0 \bigr)
=f({\rm Mn^{3+}(A)})+f({\rm Mn^{3+}(B)})-2f({\rm Mn^{4+}})+c ,
\end{equation}
with ASF of Mn$^{4+}$ $f({\rm Mn^{4+}})$ and a constant $c$. 
Both the reflections at $({2l+1 \over 4}\ {2l+1 \over 4}\ 0)$ and $({2l+1 \over 2}\ {2l+1 \over 2}\ 0)$ 
appear around 200K and show almost identical temperature dependence as shown in 
Fig.~\ref{fig:214temp}. 
This implies that the charge and orbital order parameters couple with each other. 

After the first observation, 
this technique was applied to a parent compound of CMR manganites LaMnO$_3$ \cite{murakami_113}. 
As introduced in Sec.~\ref{sec:intro}, 
an alternate ordering of orbital with a $\sqrt{2}a \times \sqrt{2} a \times 2c$
unit cell is expected from the largely distorted MnO$_6$ octahedron 
and the $A$-type AF ordering. 
The resonant behavior near the Mn $K$-absorption edge  
was observed at (300) corresponding to the orbital superlattice reflection point. 
The $\sin^2\varphi$-type azimuthal angle dependence was confirmed only at the experimental arrangement 
where the scattered x-ray polarization $\lambda_f$ is $\pi$. 
On the contrary, the intensity almost disappears in the case of $\lambda_f=\sigma$. 
This is consistent with the symmetry of the aspherical charge cloud 
shown in Fig.~\ref{fig:orbilmo3}. 
The observed temperature dependence of the RXS intensity is shown in Fig.~\ref{fig:orbilmo3t}. 
The intensity rapidly increases at 780K which coincides with the 
structural phase transition temperature from the pseudo-cubic perovskite structure ($O^\ast$ phase) to 
the orthorhombic one ($O'$ phase). 

\begin{table}[t]
\caption{A list of experimental studies of RXS}
\label{tab:table1}
\begin{center}
\begin{tabular}{c|c} 
Materials &  References \\ \hline 
LaMnO$_3$  &  \cite{murakami_113,zimmermann_2} \\
La$_{1-x}$Sr$_x$MnO$_3$  \ ($x=$0.08, 0.1, 0.11, 0.12) & \cite{endoh,fukuda,hirota_rxs}  \\
Nd$_{0.5}$Sr$_{0.5}$MnO$_3$ & \cite{noguchi} \\
Pr$_{1-x}$Ca$_{x}$MnO$_3$ ($x$=0.25, 0.4, 0.5) & \cite{zimmermann_1,zimmermann_2,zimmermann_3} \\
La$_{2-x}$Sr$_{x}$MnO$_4$ ($0.4<x<0.5$) & \cite{murakami_214,wakabayashi2} \\
LaSr$_{2}$Mn$_2$O$_7$ & \cite{chatterji,wakabayashi} \\
La$_{1-x}$Sr$_{x}$MnO$_3$ (Artificial Lattice) &  \cite{kiyama} \\
\hline
YTiO$_3$ & \cite{nakao} \\
LaTiO$_3$ & \cite{keimer} \\
YVO$_3$ & \cite{noguchi} \\
V$_2$O$_3$ & \cite{paolasini} \\
\hline
DyB$_2$C$_2$ & \cite{tanaka,hirota,matsumura} \\
CeB$_6$ & \cite{nakao2} \\
UPd$_3$ &   \cite{mcmorrow} \\ 
\end{tabular}
\end{center}
\end{table}
At the same time, independently of perovskite manganites, 
RXS was applied to V$_2$O$_3$ \cite{fabrizio,paolasini} and 
was studied theoretically and experimentally as a probe to 
detect the orbital ordering.
Nowadays, RXS has been widely utilized to study of the orbital degree of freedom in several 
manganites \cite{nakamura,endoh,zimmermann_1,fukuda,chatterji,wakabayashi,
zimmermann_2,zimmermann_3,wakabayashi2,kiyama,hirota_rxs}, 
other transition-metal oxides 
\cite{keimer,noguchi,nakao}
and $f$-electron systems \cite{tanaka,hirota,mcmorrow,nakao2,matsumura}. 
Simultaneously, theories of RXS in the several view points were developed 
\cite{ishihara_prl,ishihara_az,ishihara_layer,ishihara_group,ishihara_jmmm,ishihara_ti,
elfimov,benfatto,takahashi1,takahashi2,takahashi3,nagao,ovchinnikova,castleton,shishidou,benedetti,mahadevan}. 
This technique is also applied to detect the charge ordering in transition-metal oxides 
\cite{murakami_214,zimmermann_1,wakabayashi,inami_co,garcia,nakao_nav}.
The published experimental RXS studies for the orbital states are 
listed in table ~\ref{tab:table1}. 

\begin{figure}[t]
\begin{center}
        \resizebox{0.7\linewidth}{!}{\includegraphics{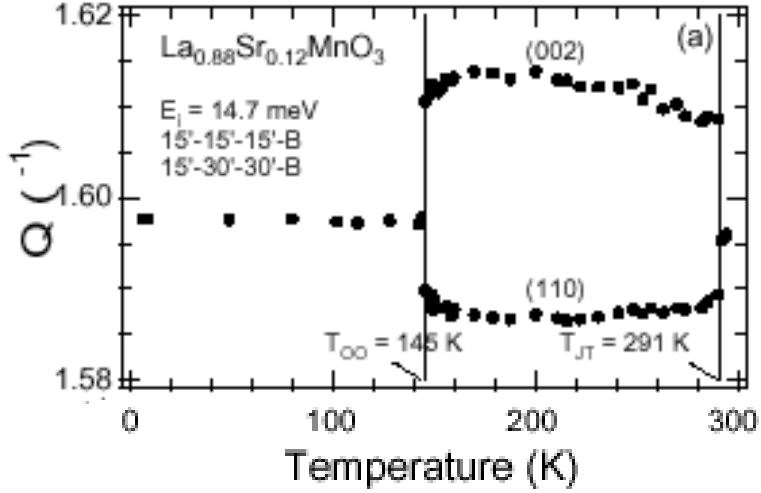}}
\end{center}
        \caption{Temperature dependence of lattice parameter in La$_{0.88}$Sr$_{0.12}$MnO$_3$ \cite{endoh}.}
        \label{fig:lts0.12}
\end{figure}

\begin{figure}[t]
\begin{center}
        \resizebox{0.75\linewidth}{!}{\includegraphics{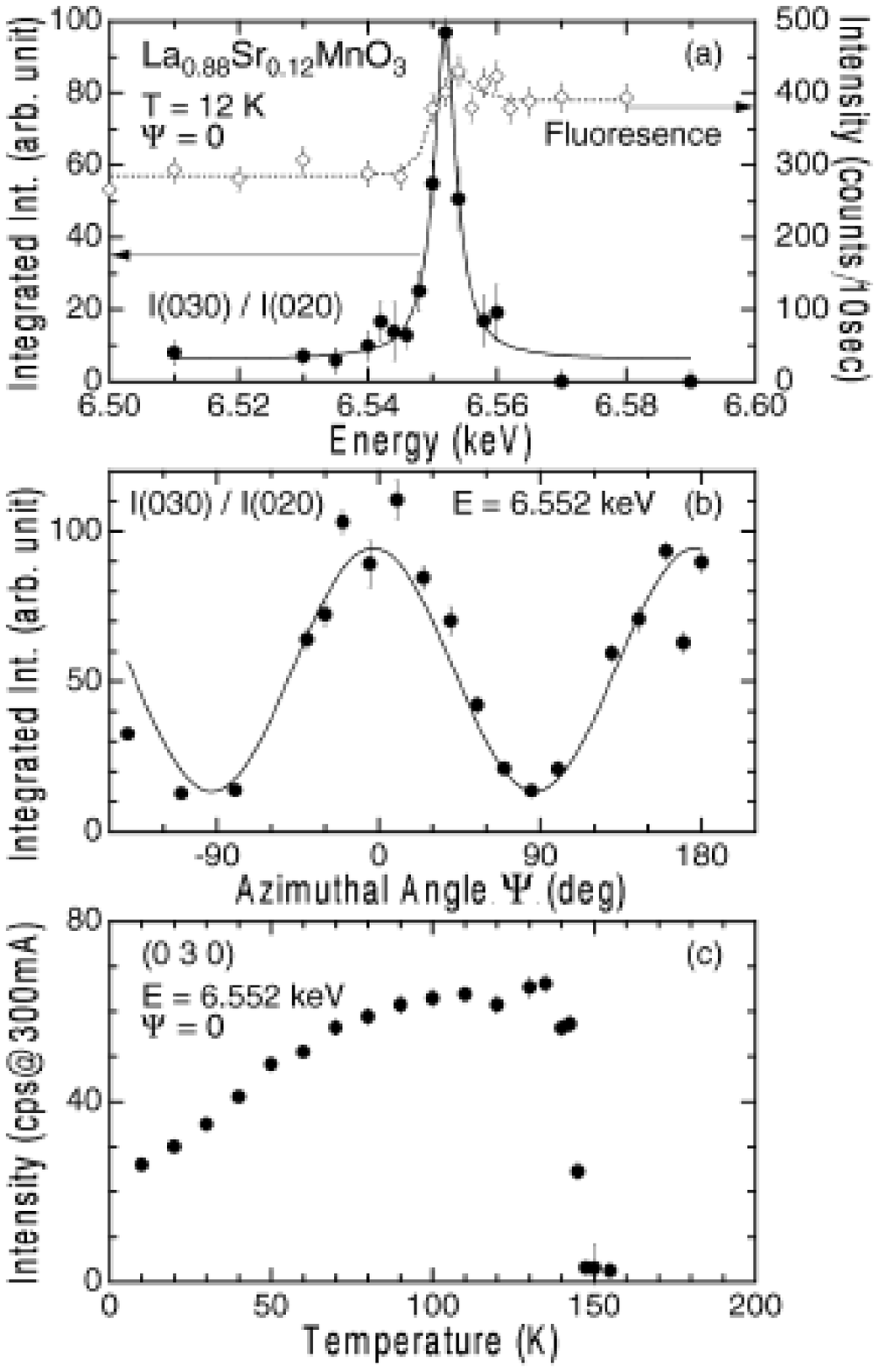}}
\end{center}
        \caption{(a) Energy dependence of RXS intensity of the 
        orbital superlattice 
        reflection (0 3 0) in La$_{0.88}$Sr$_{0.12}$MnO$_3$. The dashed curve 
        represents fluorescence. 
        (b) The azimuthal angle 
        dependence of the (0 3 0) reflection. The solid curve is the two fold squared sine curve of 
        angular dependence. 
        (c) Temperature dependence of the (0 3 0) peak intensity \cite{endoh}. }
        \label{fig:rixs0.12}
\end{figure}

\begin{figure}[t]
\begin{center}
        \resizebox{0.7\linewidth}{!}{\includegraphics{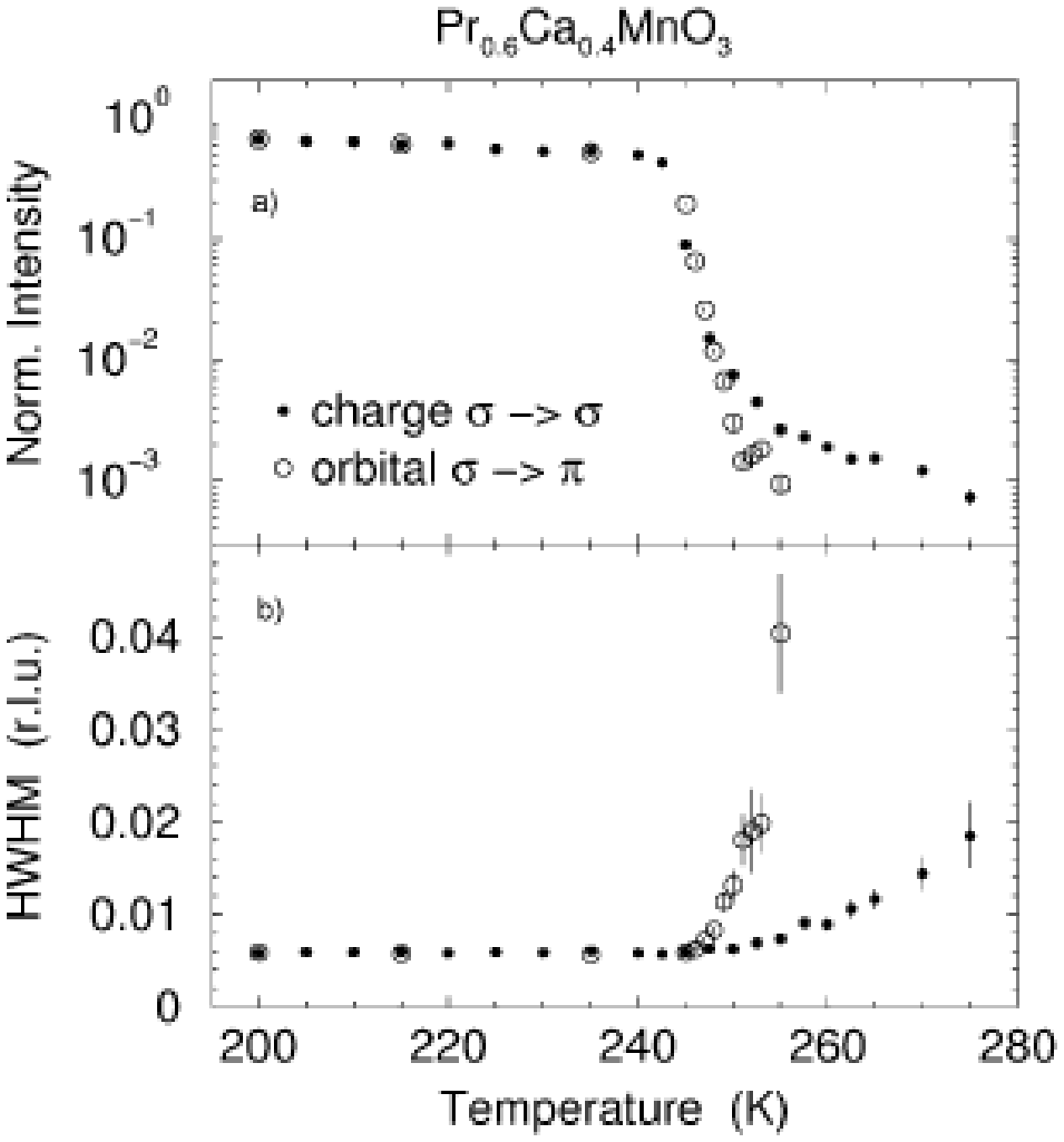}}
\end{center}
        \caption{(a) Temperature dependence of the peak intensities of 
        the (0 3 0) charge 
        superlattice reflection (closed circles) and the (0 2.5 0) 
        orbital superlattice reflection (open circles) in Pr$_{0.6}$Ca$_{0.4}$MnO$_3$. 
        (b) Temperature dependence of the half widths at half maximum \cite{zimmermann_1}. } 
        \label{fig:prca0.4} 
\end{figure}

In particular, the following two experiments provided key information for 
the mechanism of RXS and orbital ordering in manganites. 
\begin{figure}[t]
\begin{center}
        \resizebox{0.7\linewidth}{!}{\includegraphics{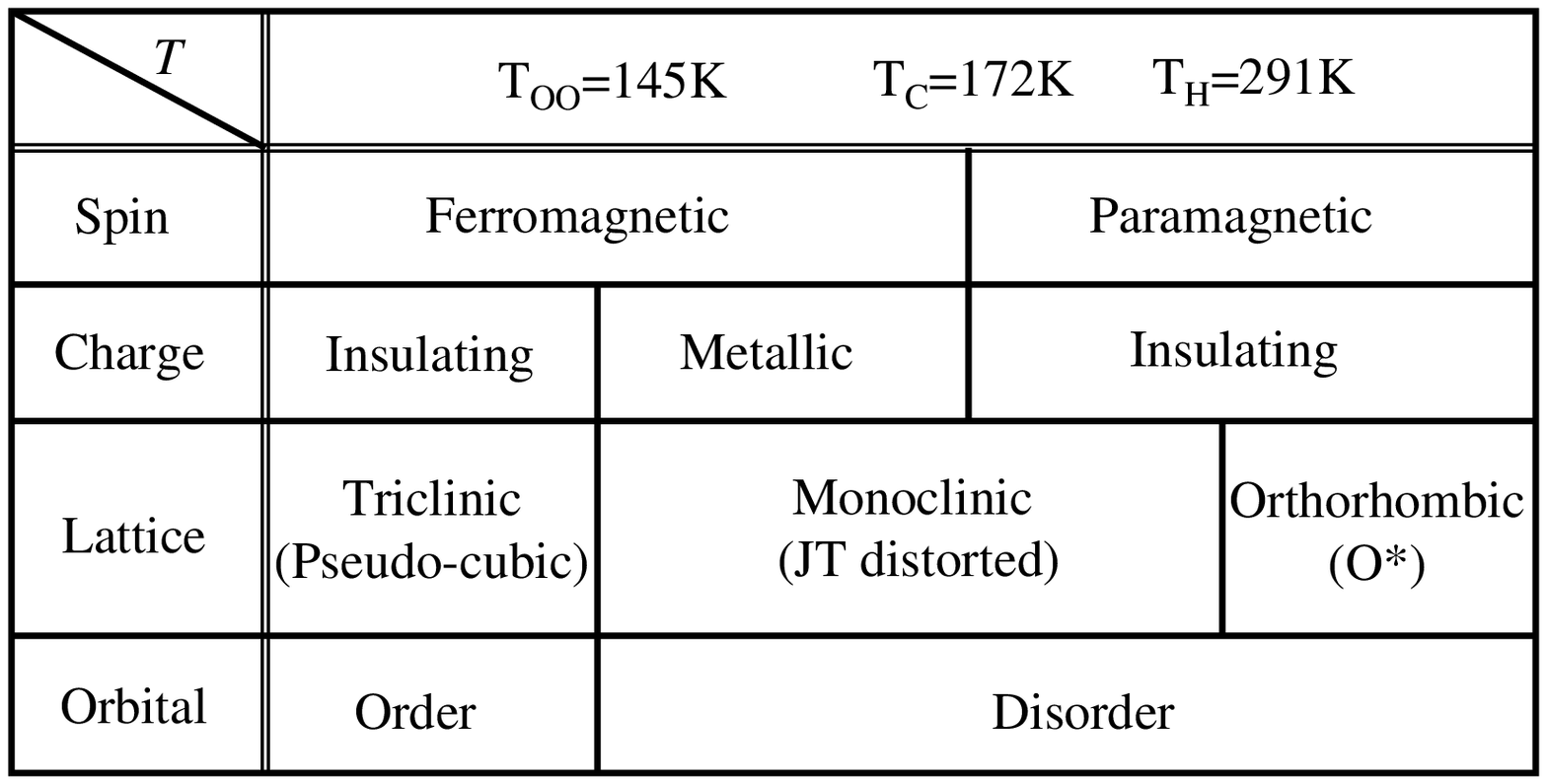}}
\end{center}
        \caption{The spin, charge, lattice 
        and orbital phase diagram  in La$_{0.88}$Sr$_{0.12}$MnO$_3$.}
        \label{fig:ps0.12}
\end{figure}
(1){\it  La$_{1-x}$Sr$_{x}$MnO$_3$ at $x \sim 1/8$} \cite{endoh,fukuda}: 
By doping of holes into LaMnO$_3$, 
the magnetic, electric and structural properties rapidly change and  
a number of phase boundaries are entangled around $x=1/8$. 
In La$_{0.88}$Sr$_{0.12}$MnO$_3$, 
the electric resistivity shows metallic behavior below the ferromagnetic transition temperature $T_C=172$K 
and shows sharp upturn below a temperature $T_{OO}=145$K \cite{tokura95,yamada}. 
Successive structural-phase transitions were observed by neutron and x-ray scattering experiments: 
the orthorhombic phase ($O^\ast$ phase) to the monoclinic one ($M$ phase ) at $T_{H}=291$K 
and the $M$ phase to the triclinic one ($T$ phase) at $T_{OO}$ \cite{kawano}. 
The Jahn-Teller type lattice distortion in MnO$_6$ octahedra, which 
is a similar type in LaMnO$_3$, 
is confirmed below $T_{H}$ but dramatically weakens at $T_{OO}$ \cite{argyriou,dabrowski,endoh}. 
The phase diagram in La$_{0.88}$Sr$_{0.12}$MnO$_3$ 
is summarized in Fig.~\ref{fig:ps0.12} and the temperature dependence of 
lattice parameter observed by the neutron scattering is presented in Fig.~\ref{fig:lts0.12} \cite{endoh}.   
RXS experiments was carried out and 
the scattering intensity was observed at (030) below $T_{OO}$ as shown in Fig.~\ref{fig:rixs0.12}. 
This is quite surprising results, since the orbital order appears in the $T$ phase where 
the Jahn-Teller type lattice distortion is almost quenched. 
This experimental result implies that (i) the present orbital ordering  
is not caused by the cooperative Jahn-Teller effects, and (ii) RXS  in this phase 
does not originate from the lattice distortion in MnO$_6$ octahedra.   
The theoretical model by taking into account the orbital order 
well explains several anomalous experiments: the magnetic field dependence of $T_{OO}$ and 
the enhancement of the magnetization at $T_{OO}$ \cite{endoh,okamoto_pt,nojiri}. 
Through the recent systematic studies of La$_{1-x}$Sr$_{x}$MnO$_3$ with a wide range of $x$ around $1/8$, 
it was found that there exists a vertical phase boundary at $x_c \sim 0.11$ \cite{cox}.  
Below $x_c$, the orbital order and the Jahn-Teller type lattice distortion simultaneously appear 
at the structural phase transition from the $O^\ast$ to $O'$ phases. 
(2) {\it  Pr$_{0.6}$Ca$_{0.4}$MnO$_3$} \cite{zimmermann_1,zimmermann_2}: 
From the structural, magnetic and transport experiments, 
it was revealed that 
the charge and orbital ordered state appears in Pr$_{1-x}$Ca$_{x}$MnO$_3$ with $0.3<x < 0.5$ \cite{tokura_pcmo}. 
The unit cells for these orderings are the same as those in La$_{0.5}$Ca$_{0.5}$MnO$_3$, 
although an excess carrier $\delta x=|x-0.5|$ may weaken the orderings. 
RXS experiments were carried out in $x$=0.4 and 0.5,  
and the reflections at $(0\ {2n+1}\ 0)$ (charge superlattice reflection) 
and at $(0\ {2n+1 \over 2}\ 0)$ (orbital superlattice reflection) were observed below 245K. 
In particular, the peak widths of both the scatterings were examined in detail, and 
the following characteristics were found out (see Fig.~\ref{fig:prca0.4}): 
(i) At low temperatures, the half of width for the half maxima (HWHM) 
of the orbital superlattice peak does not reach the experimental resolution limit, 
in contrast to the charge superlattice peak, 
and (ii)  with decreasing temperature above 245K, 
HWHM of the charge superlattice peak is more rapidly reduced than that of the orbital one. 
As mentioned later (Sec.~\ref{sec:group}), 
RXS at the orbital superlattice reflection detects 
the ordering of the $x$-component of the pseudo-spin operator $T_x$ 
which represents the tetragonal component of an electronic cloud. 
The above experiments suggest that 
the charge order fluctuation is more correlated than that of $T_x$ above the ordering temperature 
and no long range order of $T_x$ occurs in this compound. 

\section{Theoretical framework}
\label{sec:theory}

\subsection{Scattering cross section}
\label{sec:scs} 
The differential scattering cross section of x ray from electrons in solid 
is formulated by perturbational calculation  
in terms of the interactions between electrons and photons. 
We start with the Hamiltonian for the electron-photon coupled system: 
\begin{equation}
{\cal H}={\cal H}_e+{\cal H}_p+{\cal H}_{e-p} , 
\label{eq:ham}
\end{equation}
where the Hamiltonian for the electronic system ${\cal H}_{e}$, that for photons ${\cal H}_{p}$ 
and the interactions between the two ${\cal H}_{e-p}$. 
${\cal H}_{e-p}$ is divided into the two terms 
for the interactions between 
the electronic current $\vec j(\vec r)$ and the vector potential $\vec A(\vec r)$, 
and 
the electronic charge $\rho(\vec r)$ and $\vec A(\vec r)$ as follows, 
\begin{eqnarray}
{\cal H}_{e-p}&=&{\cal H}_{j}+{\cal H}_{\rho} \nonumber \\
&=&
-{e \over c}\int d \vec r \vec j(\vec r) \cdot \vec A(\vec r)
+{e^2 \over 2mc^2} \int d \vec r \rho(\vec r) \vec A(\vec r)^2. 
\end{eqnarray}
We assume that both ${\cal H}_{e}$ and ${\cal H}_{p}$ are diagonal. 
The $S$-matrix in the second order terms with respect to $\vec A(\vec r)$ 
provides the differential scattering-cross section of x ray \cite{sakurai,blume,lovesey}: 
\begin{eqnarray}
{d^2 \sigma \over d \Omega d \omega_f}&=& \sigma_T {\omega_f \over \omega_i}  
\sum_{ f } 
\Bigl |S_1 \vec e_{k_i \lambda_f} \cdot \vec e_{k_f \lambda_f}+
\sum_{\alpha \beta}  
      (\vec e_{k_i \lambda_i})_\alpha S_{2 \alpha \beta} 
      (\vec e_{k_f \lambda_f})_\beta \Bigr|^2 
\nonumber \\
&\times& \delta(\varepsilon_f+\omega
 _f-\varepsilon_i-\omega_i) , 
 \label{eq:sigma}
\end{eqnarray}
where 
\begin{equation}
S_1=\langle f | \rho(\vec k_i-\vec k_f) | i \rangle , 
\label{eq:s1}
\end{equation}
and 
\begin{eqnarray}
S_{2 \alpha \beta}&=&{m \over e^2} \sum_m \Biggl (
{ \langle f | \vec j(-\vec k_i)_{\alpha} |m \rangle 
  \langle m | \vec j( \vec k_f)_{ \beta} |i \rangle 
  \over \varepsilon_i-\varepsilon_m-\omega_f+i\eta}
  \nonumber \\
& & \ \ \ \ \ \ \  
+{\langle f | \vec j( \vec k_f)_{\beta }  | m \rangle 
 \langle m | \vec j(-\vec k_i)_{\alpha}  | i \rangle 
 \over \varepsilon_i-\varepsilon_m+\omega_i+i \eta}
 \Biggr ) .
 \label{eq:s2}  
\end{eqnarray}
This is the so-called Kramers-Heisenberg formula. 
We consider the scattering of x ray with momentum $\vec k_i$, energy $\omega_i=c|\vec k_i|$ 
and polarization $\lambda_i$, to $\vec k_f$, $\omega_f$ and  $\lambda_f$. 
$\vec e_{k_l \lambda_l}$ ($l=i,f$) is the polarization vector of x ray, and  
$| i \rangle$, $| m \rangle$ and $| f \rangle$ indicate the electronic states in the 
initial, intermediate and final scattering states  
with energies $\varepsilon_i$, $\varepsilon_m$ and $\varepsilon_f$, respectively. 
A factor: 
\begin{equation}
\sigma_T=  \biggl( {e^2 \over mc^2} \biggr )^2 , 
\end{equation} 
is the total scattering cross section of the Thomson scattering and  
$\eta$ in the denominator in $S_2$ is an infinitesimal positive constant.  
$\vec j (\vec k)$ and $\rho (\vec k)$ are defined by the Fourier transforms of $\rho(\vec r)$ and $\vec j(\vec r)$: 
\begin{equation}
\rho(\vec k) =\int d \vec r \rho(\vec r) e^{-i\vec k \cdot \vec r} , 
\end{equation}
and 
\begin{equation}
\vec j(\vec k) =\int d \vec r \vec j (\vec r ) e^{-i\vec k \cdot \vec r} ,  
\label{eq:jk}
\end{equation}
respectively. 
In the elastic scattering,  
$\rho(\vec k)$ and $\vec j(\vec k)$ are decomposed into the charge and current operators 
defined at each lattice site: 
\begin{equation}
\rho(\vec k)=\sum_l \rho_l (\vec k) e^{-i\vec k \cdot \vec r_l} , 
\end{equation}
and 
\begin{equation}
\vec j(\vec k)=\sum_l \vec j_l (\vec k) e^{-i\vec k \cdot \vec r_l} . 
\end{equation}
Then, we define the atomic scattering factor (ASF) at site $l$: 
\begin{equation}
f_{l \alpha \beta}(\vec k_i, \vec k_f) =f_{0 l}(\vec k_i, \vec k_f) \delta_{\alpha \beta}
                         +\Delta f_{l \alpha \beta}(\vec k_i, \vec k_f) , 
\label{eq:asf}
\end{equation} 
where the first and second terms are called
the normal and anomalous terms of ASF, respectively. 
These are derived from 
$S_1$and $S_{2 \alpha \beta}$ in Eqs.~(\ref{eq:s1}) and (\ref{eq:s2}) where 
$\rho(\vec k)$ and $\vec j(\vec k)$ are 
replaced by 
$\rho_l(\vec k)$ and $\vec j_l(\vec k)$, respectively. 

When the incident x-ray energy $\omega_i$ is far from  
the energy difference of the initial and intermediate electronic 
states $\Delta \varepsilon_{mi}=\varepsilon_m-\varepsilon_i$, 
$S_1$ dominates the scattering. 
On the other hand, 
when $\omega_i$ is close to $\Delta \varepsilon_{mi}$, 
the real part of the denominator in the second term of $S_2$ becomes zero and 
the scattering intensity from this term increases divergently. 
This is the resonant x-ray scattering. 
Here, a life time of the intermediate electronic state $|m \rangle$ limits 
the scattering intensity to a finite value \cite{sakurai}. 
This life time arises from the mixing of $|m \rangle$ and other 
electronic states, such as $|i \rangle$ and $|f \rangle$. 
This is not included in ${\cal H}_{e}$ in Eq.(\ref{eq:ham}), 
because it is assumed to be diagonal. 
Electron correlation effects on a core hole and/or an excited electron by x ray 
cause this effect which is represented by the self energy $\Sigma$. 
The denominator of the second term in $S_2$ is replaced by 
$\varepsilon_i-\varepsilon_m+\omega_i +i \Gamma$ 
where Re$\Sigma$ is included in the definition of $\varepsilon_m$  
and Im$\Sigma+\eta$ is denoted by $\Gamma$. 
It is known that a magnitude of $\Delta f$ at the resonant condition 
is comparable to that of $f_0$ \cite{templeton2,templeton3}. 

Here we consider that, in the intermediate scattering states, 
one electron is excited from orbital $a$ to orbital $b$  
at a same site \cite{ishihara_prl}. 
The current operator $\vec j(\vec k)$ in Eq.~(\ref{eq:jk}) 
is represented by  
\begin{equation}
\vec j(\vec k)=-{ie \over 2m} \sum_{i \sigma} e^{i \vec k \cdot \vec r_i} 
\vec A_{a b}(\vec k) 
c_{i a \sigma}^\dagger c_{i b \sigma} +H.c. , 
\label{eq:jk2}
\end{equation}
with
\begin{equation}
\vec A_{a b}(\vec k)=\int d \vec r e^{-i \vec k \cdot \vec r} 
\{       \phi_a^\ast(\vec r) \vec \nabla \phi_b(\vec r)
- \vec \nabla \phi_a^\ast(\vec r)             \phi_b(\vec r) \} . 
\label{eq:ak}
\end{equation}
$c_{i a(b) \sigma}$ is the annihilation operator of an electron with orbital $a(b)$ 
and spin $\sigma$ at site $i$, 
and $\phi_{a(b)}(\vec r)$ is the Wannier function for an electron in the orbital $a(b)$. 
$\vec r_i$ indicates the position of the $i$-th ion in a crystal lattice. 
Consider the case of the resonant scattering where 
$\omega_i$ is tuned at the $K$ edge of a transition-metal ion 
and $b$ in Eqs.~(\ref{eq:jk2}) and (\ref{eq:ak}) indicates the $1s$ orbital in this ion. 
In a crystal with inversion symmetry, 
the multipole expansion is valid. 
Note that a relevant region of the integral in Eq.~(\ref{eq:ak}) is 
determined by the average atomic radius $\langle r_{1s} \rangle$ 
of the $1s$ orbital which is much smaller than the wave length $\lambda$ of x ray. 
In the case of a Mn ion, 
the ratio $\lambda/\langle r_{1s} \rangle$ is about 1/100. 
Thus electric quadrupole transition is much smaller than the electric dipole one 
and the anomalous part of ASF, $\Delta f_{\alpha \beta}(\vec k_i, \vec k_f)$, in Eq.~(\ref{eq:asf}) 
is almost independent of the momentum of x ray, 
unlike the normal part of ASF, $f_0(\vec k_i, \vec k_f)$. 
In comparison with other experimental methods,  RXS has the following advantages to detect 
the orbital ordering in transition-metal oxides: 
(1) The wave length of x ray, which is tuned near the $K$ absorption edge of a transition-metal ion,  
is shorter than a lattice constant of the perovskite unit cell. Thus, the diffraction 
experiments can be carried out for the orbital superlattice, unlike Raman scattering and 
optical reflection/absorption experiments. 
(2) As shown in Eq.~(\ref{eq:s2}), $S_2$ is a tensor with respect to the incident and scattered 
x-ray polarizations. On the contrary, 
$S_1$ is a scalar. 
(3) By tuning the x-ray energy at the absorption edge, 
the scattering from a specific element, such as a Mn ion, can be identified. 
 
\subsection{Microscopic mechanism of RXS in orbital ordered states}
\label{sec:mechanism}

One of the main issues in RXS as a probe to detect the orbital ordering is 
its microscopic mechanism of the scattering in orbital ordered state, i.e. the 
mechanism of the anisotropic tensor elements of ASF.  
Let us consider the RXS experiments applied to the orbital ordered $3d$ transition-metal oxides. 
The initial x ray tuned around the $K$ edge of the transition-metal ion 
causes the dipole transition from the $1s$ orbital to the $4p$ orbital in the intermediate scattering states. 
Thus, this transition itself does not access to the $3d$ orbital directly  
where the orbital ordering occurs, and 
some mechanisms which bring about the anisotropy of ASF are required. 
The similar situation occurs in RXS applied to 
the quadrupolar ordering in $4f$ electron systems. 
Here, x-ray energy is tuned around the $L$ edge of $4f$ rare-earth ions 
and the $2p-5d$ transition is brought about. 
Although some methods can access directly to the orbital concerned \cite{mcmorrow,castleton,shishidou}, 
these experiments are limited as discussed later.  
This is because the wave length of x ray should satisfy both the 
resonant and diffraction conditions.  
In this section, we introduce the microscopic mechanisms of RXS 
in orbital ordered manganites. 

\begin{figure}[t]
\begin{center}
       \resizebox{0.5\linewidth}{!}{\includegraphics{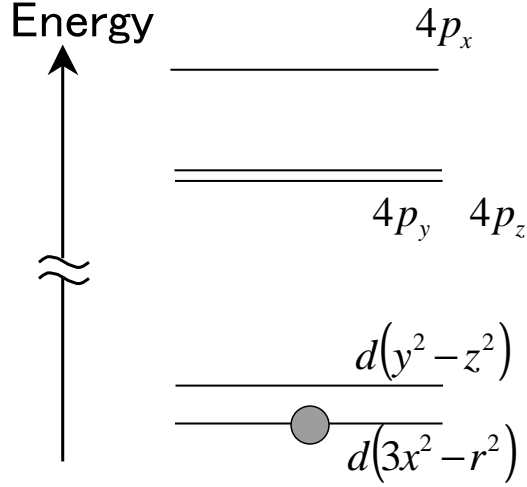}}
\end{center}
        \caption{ A schematic energy diagram of the Mn $4p$ 
        orbitals at the $d_{3x^2-r^2}$ orbital occupied site in the Coulomb mechanism.}
        \label{fig:coulomb}
\end{figure}
One of the promising candidates, which bring about the anisotropy of ASF,   
is the Coulomb interaction between $3d$ and $4p$ electrons \cite{ishihara_prl}. 
Consider an orbital ordered state in perovskite manganites. 
There exists the Coulomb interaction between $3d$ and $4p$ electrons represented by 
\begin{equation}
{\cal H}_{3d-4p}=\sum_{\gamma \gamma' \alpha \sigma \sigma'}
V_{\gamma \gamma' \alpha} 
d_{\gamma \sigma }^\dagger d_{\gamma' \sigma } 
P_{\alpha \sigma'}^\dagger P_{\alpha  \sigma'}  , 
\label{eq:h3d4p}
\end{equation}
where $d_{\gamma \sigma}$ is the annihilation operator of the $3d$ electron 
with orbital $\gamma(=3z^2-r^2,x^2-y^2)$ and spin $\sigma(=\uparrow, \downarrow)$, and 
$P_{\alpha \sigma}$ is the annihilation one of the $4p$ electron 
with the Cartesian coordinate $\alpha(=x,y,z)$. 
The Coulomb interaction $V_{\gamma \gamma' \alpha}$ between the two is given by 
\begin{eqnarray}
V_{\gamma \gamma' \alpha}=
   \left \{
\begin{array}{@{\,} ccc @{\,}}
   F_0(3d,4p)+{4 \over 35}F_2(3d,4p) \cos  (\theta_\gamma-{2 \pi \over 3} m_\alpha ) 
 & {\mbox {\rm for}} \ \gamma'=\gamma \\
   {4 \over 35}F_2(3d,4p) \sin  (\theta_\gamma-{2 \pi \over 3} m_\alpha )
 & {\mbox {\rm for}} \ \gamma'=-\gamma
 \end{array}  ,  \right . 
\label{eq:w3d4p}
\end{eqnarray}
where $(m_x, m_y, m_z)=(1,2,3)$, $-\gamma$ indicates a counterpart of $\gamma$,  
and $F_{0(2)}(3d,4p)$ is the Slater-integral between $3d$ and $4p$ electrons. 
$\theta_\gamma$ characterizes the wave funciton of 
the occupied orbital as 
$|d_\gamma \rangle =\cos(\theta_\gamma/2)| d_{3z^2-r^2} \rangle
                   -\sin(\theta_\gamma/2)| d_{x^2-y^2}  \rangle$. 
Due to this interaction, 
the three $4p$ orbitals split 
and the energy levels of the $4p_y$ and $4p_z$ orbitals relatively 
decrease at the site where the $d_{3x^2-r^2}$ orbital is occupied 
(see Fig.~\ref{fig:coulomb}). 
As a result,  
the anisotropy of ASF is brought about 
such that  
$|\Delta f_{l yy}|=|\Delta f_{l zz}|\equiv f_s$, 
$|\Delta f_{l xx}| \equiv f_l$ 
and $f_l<f_s$. 
At a site where $d_{3y^2-r^2}$ orbital is occupied, 
the conditions 
$|\Delta f_{l xx}|=|\Delta f_{l zz}|\equiv f_s$ and  
$|\Delta f_{l yy}| \equiv f_l$ is derived.  
A RXS intensity at the orbital superlattice reflection point is proportional 
to $|\Delta f_{3x^2-r^2 \alpha \alpha}-\Delta f_{3y^2-r^2 \alpha \alpha}|^2$ 
and is finite in the cases of $\alpha=x$ and $y$. 
These are consistent with the experimental results in the azimuthal angle dependence of the scattering intensity. 
In addition to the $3d-4p$ Coulomb interaction, 
the inter-site Coulomb interaction between Mn $3d$ and O $2p$ electrons brings about the anisotropy of ASF. 
Because there exists a strong hybridization between Mn $3d$ and O $2p$ orbitals, 
the $|d_{3x^2-r^2}^1 \rangle$ state strongly mixes with the 
$|d_{3x^2-r^2}^1 d_{y^2-z^2}^1{\underline L}_{y^2-z^2}\rangle$ state. 
${\underline L}_{y^2-z^2}$ indicates a state where one hole occupies the 
linear combination of the ligand O $2p$ orbitals with the 
$y^2-z^2$ symmetry. 
This interaction is given by 
\begin{equation}
{\cal H}_{2p-4p}=\sum_{\gamma \alpha \sigma \sigma'}
V_{\gamma  \alpha}^{inter} 
p_{\gamma \sigma }^{(h) \dagger} p_{\gamma  \sigma }^{(h)} 
P_{\alpha \sigma'}^\dagger P_{\alpha  \sigma'}  , 
\label{eq:h2p4p}
\end{equation}
with 
\begin{equation}
V_{\gamma \alpha}^{inter}= 
-\varepsilon - {\textstyle \frac{\varepsilon \rho^2}{5}} 
\cos (\theta_\gamma+m_\alpha {\textstyle \frac{2 \pi}{3}} ) , 
\end{equation}
where $p_{\gamma \sigma}^{(h)}$ is the annihilation operator of the $2p$ hole. 
$\varepsilon=Ze^2/a$ and $\rho=\langle r_{4p} \rangle/a$ with $Z=2$ 
and the average radius of the Mn $4p$ orbital $\langle r_{4p} \rangle$. 
The relative level structure caused by this interaction is the same with that shown in 
Fig.~\ref{fig:coulomb}. 
The two mechanisms cause the anisotropy of ASF cooperatively 
and are termed the Coulomb mechanism. 
ASF based on this mechanism were calculated in a MnO$_6$ cluster at first 
(Refs.~\cite{ishihara_prl,ishihara_az}). 
Then, the theory was developed through the calculations where the itinerant character of the $4p$ 
electrons is introduced (Refs.~\cite{ishihara_layer,ishihara_jmmm}). 
These results are reviewed in more detail later. 

\begin{figure}[t]
\begin{center}
       \resizebox{0.8\linewidth}{!}{\includegraphics{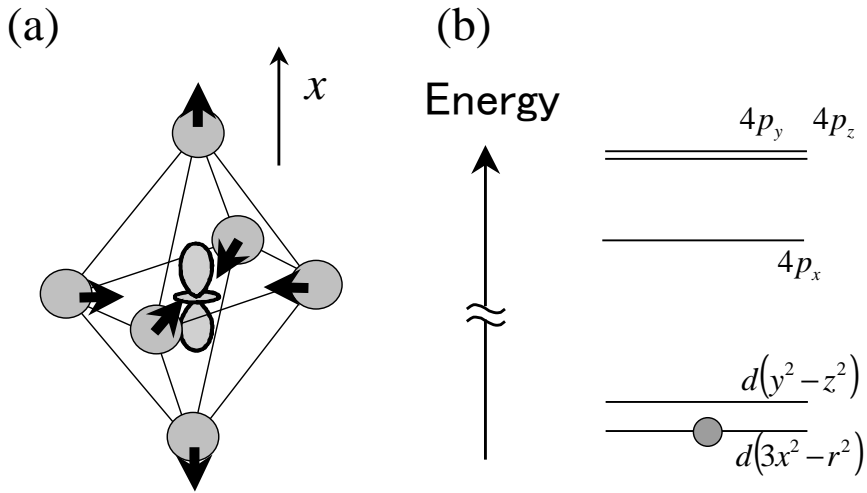}}
\end{center}
        \caption{(a) A distortion in a MnO$_6$ octahederon where 
        $d_{3x^2-r^2}$ orbital is occupied by an electron.   
        (b) A schematic energy diagram of the Mn $4p$ 
        orbitals at the $d_{3x^2-r^2}$ orbital occupied site in the Jahn-Teller mechanism.}
        \label{fig:legeljt}
\end{figure}
Another mechanism of the anisotropy of ASF was proposed from 
the view point of the Jahn-Teller type lattice distortion \cite{elfimov,benfatto,takahashi1}. 
Consider the $d_{3x^2-r^2}$ orbital occupied site.  
A lattice distortion of an O$_6$ octahedron along the $x$ direction is often observed 
around this Mn site (Fig.~\ref{fig:legeljt} (a)). 
This is termed the Jahn-Teller type lattice distortion. 
Since the Mn-O bond lengths along the $x$ directions are elongated, 
the hybridization between O $2p_x$ and Mn $4p_x$ orbitals in these bonds is weaker than others. 
As a result, 
the three $4p$ orbitals split so that the energy level of  
the $4p_x$ orbital is lower than those of the $4p_y$ and $4p_z$ orbitals (see Fig.~\ref{fig:legeljt}(b)),  
because the $4p$ orbitals are the so-called anti-bonding orbitals. 
This tendency is opposite to that due to the Coulomb mechanism  
and is termed the Jahn-Teller mechanism. 
The anisotropy of ASF at the $K$ edge is given  
such that  
$|\Delta f_{l yy}|=|\Delta f_{l zz}|\equiv f_s$, 
$|\Delta f_{l xx}| \equiv f_l$ 
and $f_l>f_s$.
This scenario was demonstrated by the LSDA+U method (Ref.~\cite{elfimov}) 
and the LDA method in the KKR scheme (Ref.~\cite{takahashi1}). 
The linear muffin-tin orbital (LMTO) method was utilized to calculate 
the local density of states (DOS) for the $4p$ electrons in the crystal structure of LaMnO$_3$.  
The calculated anisotropy of DOS was compared with the x-ray absorption and RXS spectra. 
ASF based on this mechanism was also calculated by the finite difference method and 
the multiple scattering theory in a finite cluster system (Ref.~\cite{benfatto}). 
Here the effects of the core hole potential are introduced unlike the band structure calculations. 
It is noted that the calculated spectra obtained by the two methods, i.e. 
the band calculations and the cluster calculations, show large differences 
with each other. 
This may be attributed to the itinerant effects of the $4p$ electrons 
and the core hole potential, 
which are treated properly in the former and the later, respectively. 

As mentioned above, the RXS intensity is given by $|f_{l}-f_{s}|^2$. 
This is proportional to 
the square of the level splitting of the Mn $4p$ orbitals $\Delta(\equiv \varepsilon_{x}-\varepsilon_z)$, 
where $\varepsilon_{x(z)}$ is the energy level of the Mn $4p_{x(z)}$ orbital, 
and does not depend on the sign of $\Delta$ \cite{murakami_113}. 
Then, quantitative estimations for the anisotropy of ASF are required. 
However, it is, in general, difficult theoretically to treat quantitatively 
both the local correlation effects, such as the core hole potential and the $3d$-$4p$ Coulomb interactions, 
and the large itinerant character of the $4p$ electrons on an equal footing. 
Until now, there are not direct evidences  
and the dominant mechanism of the anisotropy of ASF is still controversial. 
The measurements of the interference of $S_1$ and $S_2$ 
in Eq.~(\ref{eq:sigma}) may be available to determine the sign of $\Delta$ \cite{ishihara_az,kiyama}. 
On the other hand, 
the following experiments suggest that 
the Coulomb mechanism is expected to dominate in these compounds: 
(1) As introduced in Sect.~\ref{sec:recentex},  
the neutron scattering experiments in La$_{0.88}$Sr$_{0.12}$MnO$_3$ show that  
the Jahn-Teller type lattice distortion exists among 291K$>T>$145K, and 
this distortion is almost quenched below 145K. 
On the other hand, RXS intensity is not observed in the region of 291K$>T>$145K but 
appears below 145K \cite{endoh,fukuda}. 
That is, the region where the Jahn-Teller distortion appears 
and that where the RXS intensity is observed are different. 
These results suggest that the anisotropy of ASF 
in this compound is not caused by the Jahn-Teller mechanism. 
(2) YTiO$_3$ is known to show the orbital ordered state where 
four kinds of orbitals are arranged in a unit cell. 
RXS experiments in YTiO$_3$ were recently carried out, in detail, 
at several orbital superlattice reflection points and the polarization configurations. 
Through the quantitative analyses,  
it was concluded that the relative scattering intensities can not be explained 
by ASF expected from the lattice distortion 
but is consistent with ASF in the Coulomb mechanism \cite{nakao} . 
In order to clarify the mechanism of the anisotropy of ASF,  
the further experimental and theoretical researches are required. 

\begin{figure}[t]
\begin{center}
        \resizebox{0.75\linewidth}{!}{\includegraphics{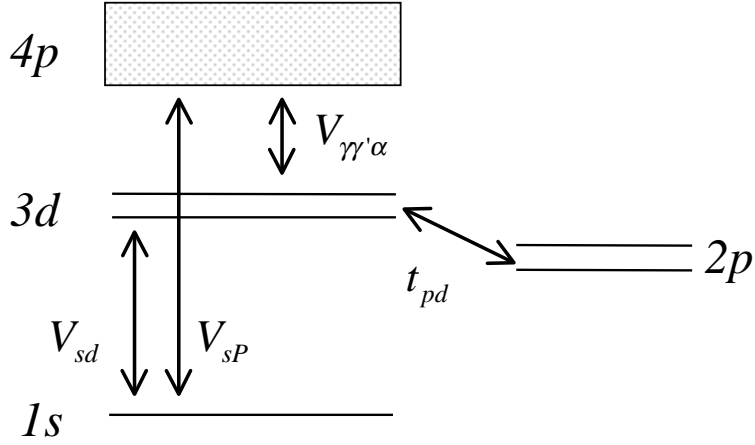}}
\end{center}
        \caption{A schematic energy diagram for the Hamiltonian in Eq.~(\ref{eq:clba}).}
\label{fig:level}
\end{figure}
By taking into account the above facts, 
we review, in the following, the calculations of ASF based on the Coulomb mechanism \cite{ishihara_jmmm,ishihara_layer}. 
Let us start with the model Hamiltonian: 
\begin{equation}
{\cal H}={\cal H}_{\rm cluster}+{\cal H}_{\rm band}. 
\label{eq:clba}
\end{equation}
${\cal H}_{\rm cluster}$ describes the electronic system in a MnO$_6$ cluster where 
x ray is absorbed. 
Electrons in this cluster couple with the Mn $4p$ band  given by ${\cal H}_{\rm band}$. 
The Mn $3d$, $4p$ and $1s$ orbitals and the O $2p$ ones 
are introduced in ${\cal H}_{\rm cluster}$, 
and the corresponding annihilation operators are defined by 
$d_{l \gamma \sigma}$, $P_{l \alpha \sigma}$ $s_{l \sigma}$ 
and $p_{l \gamma \sigma}$, respectively, with orbital $\gamma(=3z^2-r^2, x^2-y^2)$, 
spin $\sigma(=\uparrow, \downarrow)$ and Cartesian coordinate $\alpha(=x,y,z)$. 
The O $2p$ molecular orbitals are formed from linear combinations of the six O $2p$ orbitals 
in the octahedron.  
A simple cubic lattice of a Mn ion is considered in ${\cal H}_{\rm band}$ and 
the Mn $4p$ orbitals are introduced at each Mn site. 
The explicit form of the Hamiltonian is given by 
\begin{equation}
{\cal H}_{\rm cluster}={\cal H}_{3d}+{\cal H}_{3d-4p}+{\cal H}_{1s,4p}+{\cal H}_{1s-3d,4p}+{\cal H}_{2p-3d}, 
\label{eq:hcluster}
\end{equation}
with 
\begin{eqnarray}
{\cal H}_{3d}
&=&\varepsilon_d \sum_{\gamma \sigma} d^\dagger_{l \gamma \sigma} d_{l \gamma \sigma}
+U\sum_\gamma n_{l \gamma \uparrow}^d n_{l \gamma \downarrow}^d
+U'\sum_{\sigma \sigma'} n_{l \gamma \sigma}^d n_{l -\gamma \sigma'}^d \nonumber \\
&+&I\sum_{\sigma \sigma'} d^\dagger_{l \gamma \sigma} d^\dagger_{l -\gamma \sigma'} 
                             d_{l \gamma \sigma'} d_{l -\gamma \sigma} 
-J_H \vec S_l \cdot \vec S_{t l} , 
\label{eq:h3d}
\end{eqnarray}
\begin{eqnarray}
{\cal H}_{1s,4p}
   =\varepsilon_P \sum_{\alpha \sigma} P^\dagger_{l \alpha \sigma} P_{l \alpha \sigma} 
   +\varepsilon_s \sum_{\sigma}        s^\dagger_{l \sigma}        s_{l \sigma}  , 
\label{eq:h1s4p}
\end{eqnarray}
\begin{eqnarray}
{\cal H}_{1s-3d,4p}
=n_{h l} \Bigl (\sum_{\gamma } V_{sd} n^d_{l \gamma }+\sum_{\alpha }V_{sP}  n^P_{l \alpha } \Bigr)  ,  
\label{eq:h1s3d4p}
\end{eqnarray}
\begin{eqnarray}
{\cal H}_{2p-3d}
   =\varepsilon_p \sum_{\gamma \sigma} p^\dagger_{l \gamma \sigma} p_{l \gamma \sigma}
   +t_{pd} \sum_{\gamma \sigma} (d^\dagger_{l \gamma \sigma} p_{l \gamma \sigma}+H.c.) , 
\label{eq:h2p3d}
\end{eqnarray}
and 
\begin{eqnarray}
{\cal H}_{\rm band}=
\varepsilon_P \sum_{m \ne l \alpha \sigma} P^\dagger_{m \alpha \sigma} P_{m \alpha \sigma} 
+\sum_{j \delta \alpha \beta \sigma} 
t_\alpha^\beta P^\dagger_{m \alpha \sigma} P_{m+\delta_\beta \alpha \sigma} .   
\label{eq:h4p}
\end{eqnarray}
${\cal H}_{3d-4p}$ in Eq.~(\ref{eq:hcluster}) is presented in Eq.~(\ref{eq:h3d4p}), 
and a site where x ray is absorbed is denoted by $l$.  
A schematic energy level diagram is shown in Fig.~\ref{fig:level}. 
${\cal H}_{3d}$ is for the Mn $3d$ electronic system 
where 
$U$, $U'$ and $I$ are the intra-orbital Coulomb interaction, 
the inter-orbital one and the exchange interaction between $e_g$ electrons, respectively, 
and $J_H$ is the Hund coupling between $e_g$ and $t_{2g}$ spins. 
We define the number operator 
$n_{l \gamma }^d=
\sum_{\sigma }n_{l \gamma \sigma}^d=
\sum_{\sigma} d^\dagger_{l \gamma \sigma} d_{l \gamma \sigma}$ 
and the spin operator: 
\begin{equation}
\vec S_l={1 \over 2}\sum_{\sigma_1 \sigma_2 \gamma} 
d_{l \gamma \sigma_1}^\dagger (\vec \sigma)_{\sigma_1 \sigma_2} d_{l \gamma \sigma_2} , 
\end{equation}
for the $e_g$ electrons and the spin operator 
$\vec S_{t l}$ for the $t_{2g}$ electrons with $S=3/2$. 
${\cal H}_{1s-3d,4p}$ describes  
the core hole potential between a $1s$ hole 
$n^h_{l}(=2-\sum_{\sigma} s^\dagger_{l \sigma} s_{l \sigma})$
and $3d$ electrons and that between a hole and $4p$ electrons  
$n_{l \alpha}^P(=\sum_{\sigma}P_{l \alpha \sigma}^\dagger P_{l \alpha \sigma})$. 
$t_{\alpha}^\beta$ in ${\cal H}_{\rm band}$ (Eq.~(\ref{eq:h4p})) is the hopping integral between 
NN $4p_{\alpha}$ orbitals in the $\beta$ direction, 
and are parameterized as 
$t_{\alpha}^\beta=\delta_{\alpha \beta}t_\sigma^{4p}+(1-\delta_{\alpha \beta})t_\pi^{4p}$. 

The above model Hamiltonian includes a MnO$_6$ cluster where electrons couple with the Mn $4p$ band. 
Therefore, neither the conventional numerical methods in a small cluster nor the perturbational 
approaches in terms of the Coulomb interactions are utilized to calculate ASF. 
The memory-function method (the composite-operator method, the projection method) 
in the Green's function formalism \cite{ishihara_layer,forster,fulde,matsumoto}. 
is one of the unperturbational methods widely applied to study 
the electronic structure in solids. 
It is known that this method is reliable to describe the itinerant and localized 
nature of correlated electrons on an equal footing. 
The relevant part of ASF at site $l$ (Eq.~(\ref{eq:asf})) 
is represented by the Green's function as 
\begin{equation}
{\rm Im}\Delta f_{l \alpha \alpha}=-{|A_{1s 4p}|^2 \over m} 
\sum_\sigma {\rm Im}G_{l \alpha \sigma}(\omega+i\Gamma) ,  
\label{eq:green}
\end{equation}
where $G_{l \alpha \sigma}(\omega)$ is the Fourier transforms of 
\begin{equation}
G_{l \alpha \sigma}(t)=
\theta(t)\langle i | 
[ J_{l \alpha \sigma}^\dagger(t) , J_{l \alpha \sigma}(0) ] 
| i \rangle  , 
\label{eq:green2}
\end{equation}
with the operator 
$J_{l \alpha \sigma}=P^\dagger_{l \alpha \sigma } s_{l \sigma}$. 
It is convenient to introduce the relaxation function (Kubo function) at finite temperature $T$: 
\begin{eqnarray}
C_{l \alpha \sigma}(t)
&=&\theta(t)T \int^{\beta}_0 d\lambda
\langle J^\dagger_{l \alpha \sigma}(t) J_{l \alpha \sigma}(i\lambda) \rangle
\nonumber \\
&\equiv& \theta(t) \langle J_{l \alpha \sigma}^\dagger(t) J_{l \alpha \sigma}(0) \rangle_{\lambda} . 
\end{eqnarray}
There exists the relation between $G_{l \alpha \sigma}$ and $C_{l \alpha \sigma}$:  
\begin{equation}
{\rm Im} G_{l \alpha \sigma}(\omega)=\beta \omega{\rm Im}C_{l \alpha \sigma}(\omega) \Big |_{T \rightarrow 0} . 
\end{equation}
The relaxation function is calculated by utilizing 
the equations of motion method. 
The final form is 
given by the continued fraction form \cite{matsumoto}:  
\begin{equation}
C_{l \alpha \sigma}(\omega)\equiv \delta M^{(0)}(\omega) , 
\end{equation}
with 
\begin{equation}
\delta M^{(n-1)} (\omega)={I^{(n)} \over 
\omega-\bigl( M_0^{(n)}+\delta M^{(n)}(\omega) \bigr) I^{(n)-1}   }  , 
\label{eq:cont}
\end{equation}
for $n \ge 1$.  
$I^{(n)}$ is the normalization factor 
\begin{equation}
I^{(n)}=\langle \psi_n^{\dagger} \psi_n \rangle_{\lambda} , 
\end{equation}
and $M_0^{(n)}$ is the static part of the self-energy 
\begin{equation}
M_0^{(n)}=\langle (i\partial_t \psi_n^{\dagger}) \psi_n \rangle_{\lambda} , 
\end{equation} 
where $\psi_n$ is the operator product (the composite operator) defined by 
\begin{equation}
(\psi_1,\psi_2,\psi_3)=(
J_{l \alpha \sigma},
P_{l \alpha \sigma}^\dagger s_{l \sigma} \delta n_{l \gamma }^d, 
P_{l \alpha \sigma}^\dagger s_{l \sigma} \delta n^{dp}_{l \gamma }) ,
\label{eq:psi}
\end{equation} 
with 
$\delta A=A-\langle A \rangle$ and 
$n^{dp}_{l \gamma}=\sum_{\sigma }d_{l \gamma \sigma}^\dagger p_{l \gamma \sigma}$. 
These operators are treated as single quantum variables describing  
well-defined excitations in a system \cite{matsumoto}. 

Advantages of the memory functional method in this issue  
are the following: 
(1) The many-body excitations arising from the local Coulomb interactions 
in a MnO$_6$ cluster are treated by the composite operator $ \psi_n$ in 
Eq.~(\ref{eq:psi}).
For example, the operator product 
$\psi_2=P^\dagger_{i \alpha \sigma} s_{i \sigma} n_{i \gamma}^{dp}$  
describes the dipole transition from the Mn $1s$ to $4p$ orbitals associated with the 
charge transfer from O $2p$ to Mn $3d$ orbitals. 
This corresponds to the so-called well-screened intermediate state 
described by 
$|\underline{1s} \ 3d e_g^2 \ 4p^1 \ \underline L\rangle$ 
where $\underline L$ indicates that a hole occupies the ligand O $2p$ orbitals.  
The energy of this many-body excitation is lower than 
that of the so-called poor screened state $|\underline{1s} \ 3d e_g^1 \ 4p^1 \rangle$ 
where the excitation is described by the operator 
$\psi_0=P^\dagger_{i \alpha \sigma} s_{i \sigma}$. 
It is experimentally confirmed in several transition-metal oxides that
the well-screened state dominates 
the finial state of XAS and the intermediate state of RXS  
\cite{kosugi,tranquada,tolentino,sahiner95,sahiner96,croft,wu,hill_rixs}. 
By treating the operator products as single quantum variables, 
the many-body excitations and the configuration interactions 
between them are taken into account in this formalism.  
These are important 
on the calculation of ASF 
and are not taken into account by the independent 
single-particle scheme with an averaged potential, 
such as the LDA band calculation \cite{elfimov,takahashi1}. 
(2) 
The itinerant effects of the $4p$ electrons are included in 
$\delta M^{(n)}$. 
By applying the loop approximation in the diagramatic technique to  
some terms in $\delta M^{(n)}$, 
the itinerant character of the excited electron is taken into account. 
This effect is of crucial importance to describe the energy dependent ASF, 
since the band width of the Mn $4p$ electron is of the order of 10$eV$. 
This is not well described by the calculations 
in a small size cluster \cite{ishihara_prl,benfatto}.  

\begin{figure}[t]
\begin{center}
        \resizebox{0.75\linewidth}{!}{\includegraphics{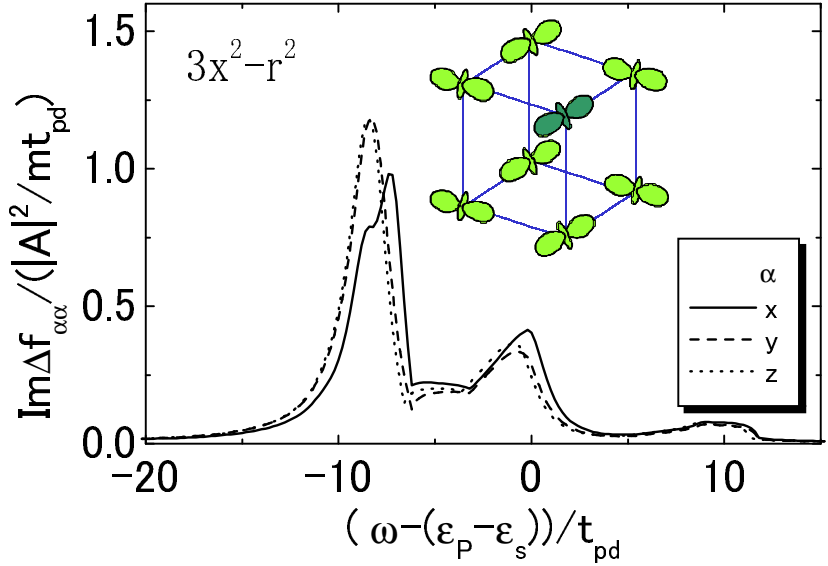}}
\end{center}
        \caption{The energy dependence of the imaginary part of ASF.
        The $3d_{3x^2-r^2}$ orbital is occupied in the 
        $[3d_{3x^2-r^2}/3d_{3y^2-r^2}]$-type orbital ordered state.
        The full, dashed and dotted lines indicate ASF for $\alpha=$
        $x$,$y$ and $z$, respectively. The inset shows a schematic picture 
        of this orbital ordered state \cite{ishihara_jmmm}.    
}\label{fig:asf}
\end{figure}
\begin{figure}[t]
\begin{center}
        \resizebox{0.75\linewidth}{!}{\includegraphics{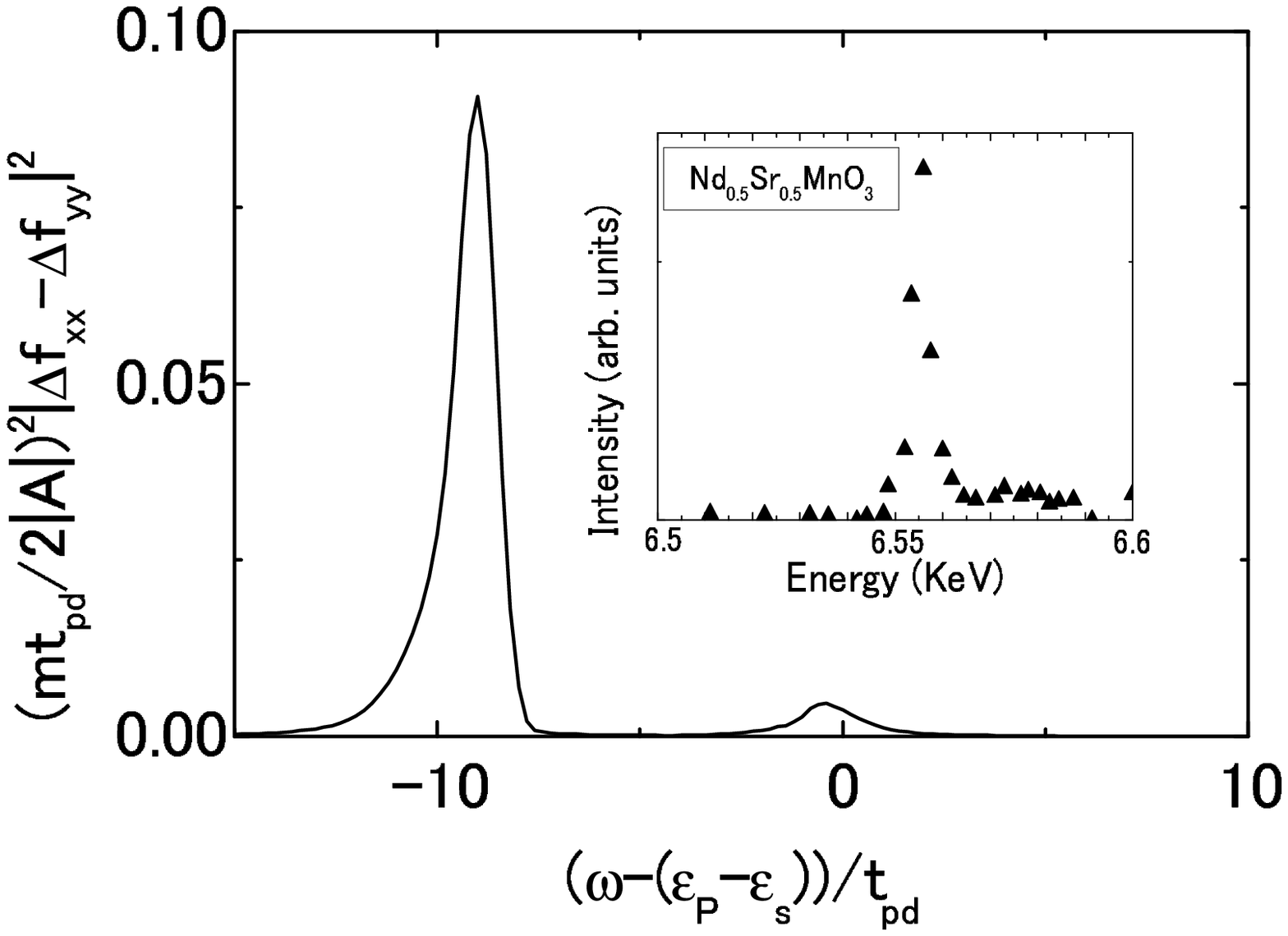}}
\end{center}
        \caption{The energy dependence of the scattering intensity of RXS  
        in the $[3d_{3x^2-r^2}/3d_{3y^2-r^2}]$-type orbital ordered state \cite{ishihara_jmmm}.
        The inset is the experimental results of the scattering intensity in 
        Na$_{0.5}$Sr$_{0.5}$MnO$_3$ \cite{nakamura}. 
        }\label{fig:intensity}
\end{figure}
The calculated ASF in RXS 
is presented in Fig.~\ref{fig:asf} \cite{ishihara_layer,ishihara_jmmm}. 
The $[3d_{3x^2-r^2}/3d_{3y^2-r^2}]$-type orbital ordered state, 
where the $3d_{3x^2-r^2}$ and $3d_{3y^2-r^2}$ orbitals are 
alternately ordered in the $ab$ plane in a cubic crystal, 
are considered and ASF at the $3d_{3x^2-r^2}$ orbital occupied site is shown. 
The parameter values in the Hamiltonian 
are chosen to be $\varepsilon_p-\varepsilon_d=1.5$, $t_{\sigma}^{4p}=3.5$, 
$t_\pi^{4p}=t_\sigma^{4p}/4$, 
$U=5$, $U'=4$, $J_H=I=1$, $V_{sd}=-5$, $V_{sp}=-3.5$,
$F_0(3d,4p)=3.5$, $F_2(3d,4p)=0.25$ and $\Gamma=1.5$ 
in units of $t_{pd}$ being about 1$\sim$1.5 eV.  
Some of these values are determined by considering 
the analyses of the photoemission and x-ray absorption 
experiments (Refs.~\cite{saitoh,park,chainani,tranquada}). 
It is interpreted that the present value of $F_2(3d,4p)$ includes the effects of 
the inter-site Coulomb interaction (Eq.~(\ref{eq:h2p4p})) 
which is not considered explicitly in the model Hamiltonian (Eq.~(\ref{eq:hcluster})). 
This is because 
(1) these two interactions contribute to the anisotropy of ASF cooperatively, 
and (2) the inter-site Coulomb interaction has a large contributions 
in the well screened state dominating the intermediate state of RXS. 
In Fig.~\ref{fig:asf}, 
the lower edge of the spectra around $(\omega-(\varepsilon_P-\varepsilon_s))/t_{pd}=-10$  
corresponds to the $K$ absorption edge of the Mn ion. 
Im$\Delta f_{l \alpha \alpha}$ shows a continuous spectra which spread over a wide region 
of energy $\omega$. 
The spectra reflect a broad width of the Mn $4p$ band. 
However, the energy dependence of Im$\Delta f_{l \alpha \alpha}$ is not the $4p$ density of states itself; 
there exist several peak structures 
in Im$\Delta f_{l \alpha \alpha}$ which originate from   
the local excitations in the MnO$_6$ cluster.
Near the absorption edge,  
the anisotropy between Im$\Delta f_{l xx}$ and Im$\Delta f_{l yy(zz)}$ is 
shown and Im$\Delta f_{l yy(zz)}$ governs the intensity. 
This anisotropy is caused by the Coulomb interaction between $3d$ and $4p$ electrons 
which pushes up the $4p_x$ band at the $3d_{3x^2-r^2}$ orbital occupied site.  
The core hole potential makes the anisotropy remarkable, 
since the potential reduces the energy of the dipole transition and 
enhances the local character of the $4p$ electrons. 
The scattering intensity of RXS,  
$I=(mt_{pd}/2|A_{1s4p}|)^2|(\Delta f_{l xx}-\Delta f_{l yy})|^2$, 
is shown in Fig.~\ref{fig:intensity} associated with 
the experimental data in Nd$_{0.55}$Sr$_{0.45}$MnO$_3$ \cite{nakamura}.  
A sharp peak structure near the $K$ edge together with 
a small intensity above the edge appears in ASF. 
Both structures were confirmed in the experimental spectra 
in several manganites with orbital order \cite{murakami_113,zimmermann_1,wakabayashi}. 

In the last part of this subsection, 
we review other theoretical examinations of RXS in orbital ordered state. 
In several cases of the orbital ordering, 
the direct access of RXS to the orbital concerned were examined. 
The orbital ordering in La$_{1.5}$Sr$_{1.5}$MnO$_4$ 
has a unit cell with $\sqrt{2}a \times 2\sqrt{2} a \times c $ (Fig.~\ref{fig:coorder}). 
As pointed out in Ref.~\cite{castleton}, 
it is barely possible to satisfy the diffraction condition 
for RXS at the Mn $L$ edge;  
$2\sqrt{2}a \sim 11 \AA$ and the wave length of x ray is about 20$\AA$.
Here, an electron is directly excited to the Mn $3d$ orbitals from the Mn $2p$ ones. 
Another possibility was proposed in the orbital ordered system 
where the local inversion symmetry is broken at the transition metal site. 
The unoccupied $3d$ states mix with the $4p$ states which are directly accessed 
by RXS at the $K$ edge. 
The experimental results of RXS in V$_2$O$_3$ \cite{paolasini} 
were compared with the theoretical calculations 
of the possible orbital ordered states (Refs.~\cite{fabrizio,ezhov,mila,natoli}). 
RXS at the $M$ edge of a $5f$ ion is also possible to directly 
access to the $5f$ orbital by the transition of $3d \rightarrow 5f$. 
RXS experiments was performed in UPd$_3$ \cite{mcmorrow}  where  
x ray with the wave length of 3.5$\AA$ was utilized. 
This process was proposed theoretically in URu$_2$Si$_2$ \cite{shishidou}. 
RXS in the orbital ordered state was also examined 
from the phenomenological point of view in Ref.~\cite{ovchinnikova}. 
The x-ray susceptibiliy tensor 
was formulated in the system where 
the several anisotropic factors due to the spin, orbital and lattice degrees of freedom coexist.  
The forbidden reflections caused by the combined effects among them 
were pointed out. 
This approach may have relation to the theoretical framework of RXS presented in Sec.~\ref{sec:group}. 

\subsection{Azimuthal angle dependence}
\label{sec:azimuth}

Once the matrix elements of ASF tensor are obtained microscopically, 
the scattering intensity in RXS is calculated in a realistic experimental arrangement. 
One of the important experimental parameters is the azimuthal angle ($\varphi$) 
which is a rotation angle in terms of the scattering vector. 
As mentioned previously, the tensor character of ASF due to the orbital order 
is directly observed through the $\varphi$ dependent scattering intensity. 
In this subsection, 
the general formulas of the scattering intensity by taking into account the 
experimental arrangements are derived and a validity of 
the $\varphi$ dependent scattering intensity to study the orbital ordering is demonstrated. 
Experimental results observed in layered manganites La$_{2-2x}$Sr$_{1+2x}$Mn$_2$O$_7$ 
are also reviewed. 

\begin{figure}[t]
\begin{center}
        \resizebox{0.8\linewidth}{!}{\includegraphics{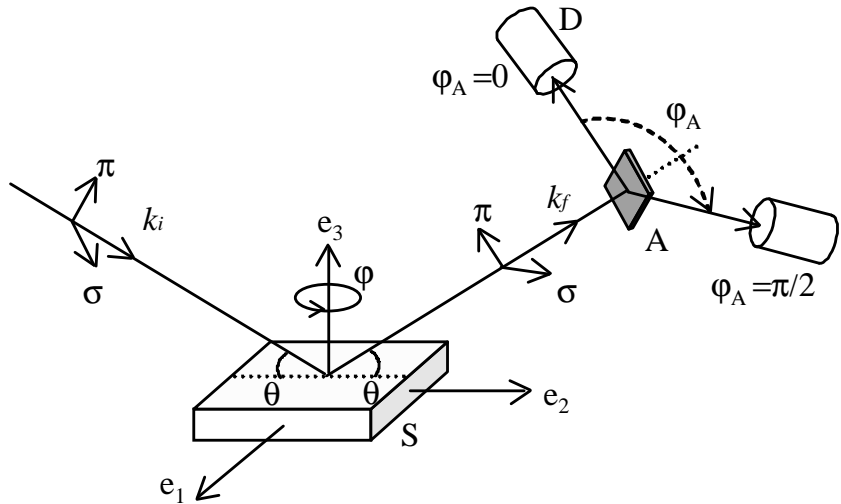}}
\end{center}        
\caption{An experimental arrangement of RXS consisting 
        of a sample crystal ($S$),  an analyzer crystal ($A$) and a photon detector ($D$).
        $\vec k_i$ and $\vec k_f$ are the incident and scattered x-ray momenta, 
        $\varphi$ is the 
        azimuthal angle, $\varphi_A$ is the analyzer angle and $\theta_s$ is 
        the scattering angle \cite{nagano,ishihara_az}. }
\label{fig:arrange}
\end{figure}
Consider an experimental arrangement of RXS applied to orbital ordered compounds 
(Fig.~\ref{fig:arrange} \cite{nagano}). 
It consists of a sample crystal $(S)$, a polarization analyzer 
including an analyzer crystal $(A)$, and a photon detector $(D)$. 
The polarization scan is characterized by two rotation angles, i.e. the azimuthal 
angle $(\varphi)$ and the analyzer angle ($\varphi_A$). 
The former is the rotation angle of the sample around the scattering vector 
$\vec K=\vec k_i- \vec k_f$, and the latter is that of the analyzer around an axis 
which is parallel to the scattered photon momentum. 
We assume that the incident x ray is perfectly polarized 
in the horizontal plane, i.e. $\sigma$-polarization. 
A direction of the electric vector of the incident x ray with respect to 
the crystalline axis is changed by the azimuthal rotation. 
Because of the tensor character of ASF, 
the scattered x ray has both the $\pi$- and $\sigma$-polarized components, 
which are separated by the analyzer scan. 
In this optical arrangement, 
the scattering intensity is given by \cite{ishihara_az}
\begin{eqnarray}
I(\varphi,\varphi_A)=\sum_{\lambda}
\bigl |
\sum_{\lambda_f} M_{\lambda \lambda_f}(\varphi_A) A_{\lambda_f \lambda_i}(\varphi)
\bigr |^2 ,  
\label{eq:intensity}
\end {eqnarray}
where $\lambda$ and $\lambda_{i(f)}$ $(=\pi, \sigma)$ indicate the polarization of x ray . 
$M_{\lambda' \lambda}(\varphi_A)$ is the scattering matrix of the analyzer:  
\begin{equation}
 M(\varphi_A)= F_A
\pmatrix{ \cos \varphi_A & -\sin \varphi_A  \nonumber \cr
         \sin \varphi_A \cos 2 \theta_A & \cos \varphi_A \cos 2 \theta_A \cr }  , 
\label{eq:analyser}
\end {equation}
with the scattering factor $|F_A|$ and 
the scattering angle $\theta_A$ in the analyzer crystal. 
For simplicity, $\theta_A$ is fixed to be $\pi/4$ from now on. 
The $\sigma$- and $\pi$-polarized components in the scattered x ray are 
separately detected by a detector with $\varphi_A=0$ and $\pi/2$, respectively. 
The scattering amplitude 
$A_{\lambda_f \lambda_i}(\varphi)$
is defined by 
\begin{eqnarray}
A_{\lambda_f \lambda_i}(\varphi)&=&\frac{e^2}{mc^2} 
\vec e_{k_f \lambda_f}
\biggl [
U(\varphi) V 
F
 V^\dagger U(\varphi)^\dagger 
\biggr ] 
\vec e_{k_i \lambda_i}^{\ t} ,  
\label{eq:amplitude}
\end {eqnarray}
where 
$F_{\alpha \beta}$ is the structure factor  
given in the coordinate of the crystallographic axis $(\hat a, \hat b, \hat c)$. 
The polarization vectors of incident and scattered x ray are given by 
\begin{eqnarray}
 \vec e_{k_i \sigma}&=&\pmatrix{1, & 0, & 0 } , \nonumber \\
 \vec e_{k_f \sigma}&=&\pmatrix{1, & 0, & 0 } , \nonumber \\
 \vec e_{k_i \pi}&=&\pmatrix{\sin \theta, & \cos \theta, & 0 } , \nonumber \\
 \vec e_{k_f \pi}&=&\pmatrix{-\sin \theta, & \cos \theta, & 0 } .
\end{eqnarray}
The unitary matrix $ U(\varphi) $ in Eq. (\ref{eq:amplitude}) describes the azimuthal rotation 
of the sample around the $\hat e_3$ axis and 
the matrix $V$ governs the transformation from 
a coordinate of the crystallographic axis $(\hat{a},\hat{b},\hat{c})$ 
to the laboratory system 
$(\hat{e_1},\hat{e_2},\hat{e_3})$. 
$F_{\alpha \beta}$ 
is the structure factor given by 
\begin{equation}
{F}_{\alpha \beta}=N\sum_{l \in cell} 
e^{i(\vec k_i-\vec k_f) \cdot \vec{r_l}} f_{l \alpha \beta}  \ , 
\label{eq:structure}
\end {equation} 
with the number of unit cell $N$. 
When the matrix $V$ and the tensor elements of ASF  
$f_{l \alpha \beta}$ are identified, 
the scattering intensity is calculated as a function of 
the azimuthal angle $\varphi$. 

\begin{figure}[t]
\begin{center}
        \resizebox{0.7\linewidth}{!}{\includegraphics{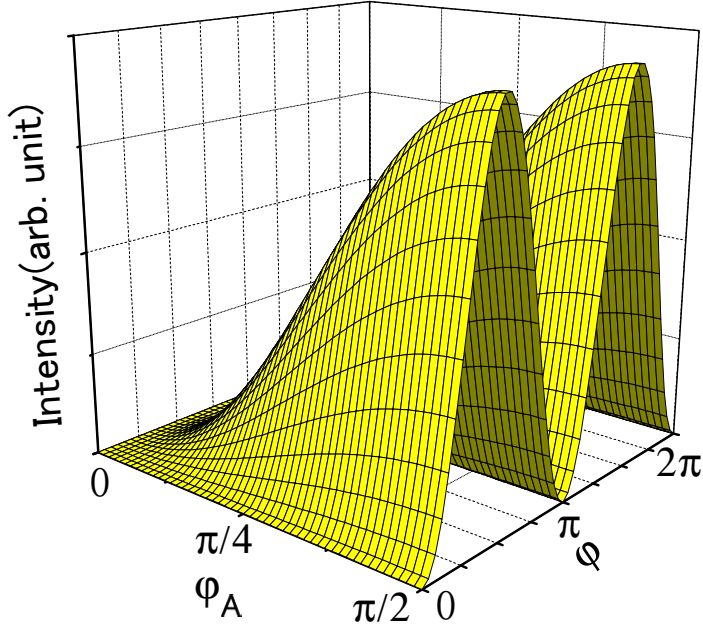}}
\end{center}
        \caption{The azimuthal and analyzer angle dependences 
        of the RXS intensity at the 
        orbital superlattice reflection \cite{ishihara_az}. \label{fig:azimuthal}}
\end{figure}
We introduce the following three cases 
where the above formulas are applied to the orbital ordered states \cite{ishihara_az}. 
(1) {\it the orbital superlattice reflection}: 
Consider the orbital ordered state where two different kinds of orbitals 
are ordered in a simple cubic lattice.
This state is termed antiferro(AF)-type orbital ordered state, on the analogy of  
the antiferromagnetic state.  
In particular, the $C$-type AF orbital ordered state, 
where two different orbitals are alternately aligned in the $ab$ plane,  
is widely observed in manganites. 
RXS at $(h \ k \ l)=({2l+1 \over 2} {2m+1 \over 2} n)$ 
termed the orbital superlattice reflection, 
appears in this state. 
Consider the case $l=m$.  
The explicit form of the scattering intensity $\tilde I (\varphi,\varphi_A)$
at this reflection point 
normalized by a factor $N^2\sigma_T|F_A|^2$ 
is obtained as: 
\begin{eqnarray}
\tilde {I} (\varphi,\varphi_A=0)&=& | \Delta f_{-zz}  \cos^2 \varphi
+{\textstyle \frac{1}{2}}( \Delta f_{-xx}+\Delta f_{-yy} ) \sin^2 \varphi |^2  , 
\label{eq:forbid10}
\end{eqnarray}
and 
\begin{eqnarray}
\tilde {I} (\varphi,\varphi_A = {\textstyle \frac{\pi}{2}} )&=&
| {\textstyle \frac{1}{2}} \bigl( -\Delta f_{-xx}+\Delta f_{-yy} ) \sin \varphi \cos \theta 
\nonumber \\
 &+& {\textstyle \frac{1}{2}} \bigl(\Delta f_{-xx}+\Delta f_{-yy}-2\Delta f_{-zz}  )  
\sin \varphi \cos \varphi \sin \theta  |^2  , 
\label{eq:forbid11}
\end{eqnarray}
which correspond to the scattering intensity with $\lambda_i=\lambda_f=\sigma$ 
($\sigma-\sigma$ scattering), and that with $\lambda_i=\sigma$ and  
$\lambda_f=\pi$ ($\sigma-\pi$ scattering), respectively. 
$ \Delta f_{- \alpha \alpha}$ is termed the AF component of ASF defined by  
$ \Delta f_{- \alpha \alpha}=\frac{1}{2} (\Delta f_{A \alpha \alpha}- \Delta f_{B \alpha \alpha})$ 
where $\Delta f_{A(B)}$ is the anomalous part of 
ASF for the $A(B)$ orbital sublattice. 
In particular, 
consider the $[\theta_A/\theta_B=-\theta_A]$-type orbital ordered state 
where the following conditions are satisfied;  
$\Delta f_{Axx}=\Delta f_{Byy}$, 
$\Delta f_{Ayy}=\Delta f_{Bxx}$ and 
$\Delta f_{Azz}=\Delta f_{Bzz}$. 
The intensity is simplified as  
\begin{equation}
\tilde {I} (\varphi,\varphi_A =0)=0 , 
\label{eq:zero}
\end{equation}
and 
\begin{equation}
\tilde {I} (\varphi,\varphi_A ={\textstyle \frac{\pi}{2}})=
 \bigl | \Delta f_{-xx} \bigr |^2 
\sin^2 \varphi \cos^2 \theta .
\label{eq:sin2}
\end{equation} 
A whole feature of $\tilde I(\varphi,\varphi_A)$ is shown in Fig.~\ref{fig:azimuthal}. 
$\tilde {I} (\varphi,\varphi_A =\pi/2)$ shows a square of the sinusoidal curve.  
The intensity becomes its maximum (minimum) at $\varphi=(2n+1)\pi/2$ ($\varphi=n\pi$) 
where the electric field of incident x ray is parallel to the $ab$ plane (the $c$ axis). 
This is explained by the fact that 
the AF component of ASF  
$ \Delta f_{- \alpha \alpha}$ 
is zero for $\alpha=z$ and is finite for $\alpha=x$ and $y$. 
Note that the $\varphi$ dependence of the scattering intensity 
in other types of orbital ordered state is distinct from the above results. 
It has been experimentally confirmed, in several perovskite manganites, that  
the observed $\varphi$ dependence  
at the orbital superlattice reflection is well fitted 
by Eqs.~(\ref{eq:zero}) and (\ref{eq:sin2}).  
Then, it is concluded that the orbital ordered state in these compounds to be
of the $[\theta_A/-\theta_A]$-type, although a value of $\theta_A$ is not determined. 

As shown above, the AF component of ASF, $\Delta f_{- \alpha \alpha}$, is observed 
at the orbital superlattice reflection, because a phase of x ray at the orbital sublattice $A$ 
is different from that at $B$ by $\pi$. 
On the other hand, 
the ferro(F)-component of ASF defined by 
$ \Delta f_{+ \alpha \alpha}=\frac{1}{2} 
(\Delta f_{A \alpha \alpha}+\Delta f_{B \alpha \alpha})$  
is not detected at this reflection. 
The F component of ASF 
reflects from the uniform shape of orbital.  
A value of $\theta_A$ in the $[\theta_A/\theta_B=-\theta_A]$-type orbital ordered state 
can be determined by $\Delta f_{+ \alpha \alpha}$. 
We introduce, in the following, the two possible methods by which 
the F component of ASF is observed. 
(2){\it the fundamental reflection}: 
The F component of ASF directly reflects the scattering intensity 
at the fundamental reflection point denoted by 
$(h \ k \ l)=(l\ m\ n)$.
The scattering intensity at this point is given by 
\begin{eqnarray}
\tilde {I} (\varphi,\varphi_A=0)=
|\Delta f_{+xx} \cos^2 \varphi+\Delta f_{+zz} \sin^2 \varphi+f_{0+}|^2 ,  
\label{eq:ffund4} 
\end {eqnarray} 
where $f_{0+}$ is defined by $f_{0+}={1 \over 2} (f_{0A}+f_{0B})$. 
$f_{0+}$ is a scalar and 
does not show the $\varphi$ dependence. 
Thus, the F component of ASF, $\Delta f_{+ \alpha \alpha}$, is observed by the azimuthal scan.  
This is an analogous to the fact that 
the ferromagnetic component is obtained through the polarization analyses of the neutron scattering.  
Since, even in the resonant scattering, 
the scattering intensity from $f_0$ is usually larger than that from $\Delta f_{+ \alpha \alpha}$,  
the interference effects between the normal and anomalous parts of ASF in Eq.~(\ref{eq:ffund4})
dominates the $\varphi$ dependence. 
In an actual experiment, a higher order reflection is effective, 
since $f_0$ decreases rapidly with $|K|$ in contrast to $\Delta f_{+ \alpha \alpha}$. 
A small tip of the azimuth rotation axis from the crystarographic axis  
is useful to measure this interference effect, 
as recently performed in some manganites \cite{kiyama}. 
(3){\it the charge order superlattice reflection}: 
When the orbital order appears associated with the charge order, 
the F component of ASF is obtained by utilizing the 
$\varphi$ dependent RXS intensity at the charge superlattice reflection. 
Consider the charge and orbital ordered state 
observed in manganites around  hole concentration being $0.5$ 
(see Fig.~\ref{fig:coorder}). 
The charge order is observed by RXS at  
$(h \ k \ l)=({2l+1 \over 2} \ {2m+1 \over 2} \ n)$ 
termed the charge superlattice reflection. 
In the case of $l=m$, 
the scattering intensity is given by 
\begin{eqnarray}
\tilde {I} (\varphi,\varphi_A=0)= {\textstyle \frac{1}{4}}
|\Delta f_{+xx}\sin^2 \varphi + \Delta f_{+zz}\cos^2 \varphi
-\Delta f_4|^2 ,
\label{eq:coo}
\end{eqnarray}
where 
$\Delta f_4$ is the anomalous part of ASF for Mn$^{4+}$ 
which is independent of the polarization of x ray. 
Thus, the interference term of $\Delta f_4$ and $\Delta f_{+ \alpha \alpha}$, 
as well as a term $|\Delta f_{+ \alpha \alpha}|^2$, 
gives rise to the $\varphi$ dependent intensity. 
A much remarkable $\varphi$ dependence is expected 
in comparison with that in the case $(2)$ \cite{ishihara_az,nakamura}. 

\begin{figure}[t]
\begin{center}
        \resizebox{0.75\linewidth}{!}{\includegraphics{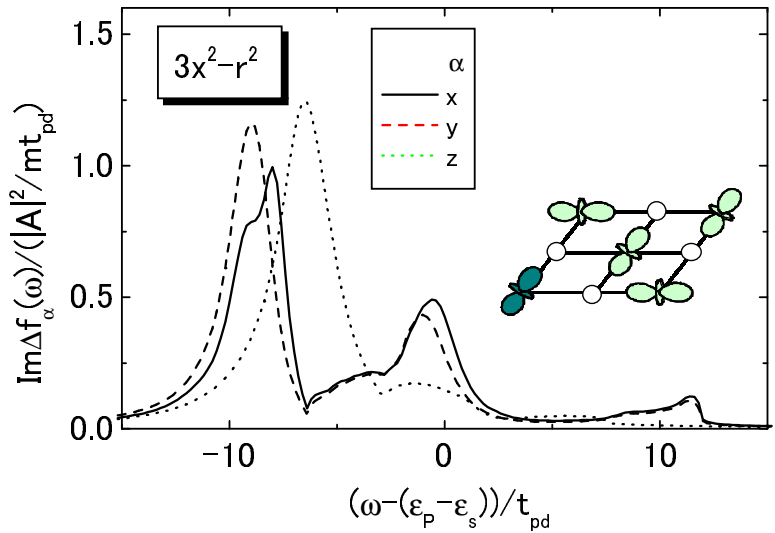}}
\end{center}
        \caption{The energy dependence of the imaginary part of ASF 
        in the layered manganites. The $3d_{3x^2-r^2}$ orbital is occupied in the 
        $[3d_{3x^2-r^2}/3d_{3y^2-r^2}]$-type orbital ordered state.
        The full, dashed and dotted lines indicate ASF for $\alpha=$
        $x$,$y$ and $z$, respectively. Inset shows a schematic picture of the 
        charge and orbital ordered state \cite{ishihara_layer}.}
        \label{fig:asflayer}
\end{figure}
So far, assuming that a crystal lattice has the cubic symmetry,   
the azimuthal angle dependence of RXS have been studied. 
Now we introduce another example of the $\varphi$ dependent scattering intensity 
from the charge and orbital orderings in a 
layered manganites La$_{2-2x}$Sr$_{1+2x}$Mn$_2$O$_7$ \cite{wakabayashi,ishihara_layer}. 
This compound is recognized to be a material appropriate 
for studying orbital degree of freedom \cite{moritomo_327,kimura,kubota,
medarde,ling,ishihara_layer2,okamoto_327,hirota_327,dessau}, 
because the spin and orbital states are systematically changed with 
the hole concentration 
In LaSr$_{2}$Mn$_2$O$_7$ ($x=0.5$), 
the charge and orbital ordering appears between about 100K and 210K \cite{wakabayashi}. 
Since a MnO$_6$ octahedron in this compound is almost isotropic, 
the energy levels of the two $e_g$ orbitals as well as the three $4p$ orbitals  
are nearly degenerate in the local sense. 
On the other hand, the layered structure 
provides the quasi-two dimensional character of the $4p$ band 
and lifts the degeneracy. 
Therefore, both the layered structure as well as the orbital ordering 
are crucially important on the anisotropy in ASF. 
ASF in this compound was formulated by 
utilizing the memory function method introduced in Sec.~\ref{sec:mechanism}. 
\begin{figure}[t]
\begin{center}
        \resizebox{0.75\linewidth}{!}{\includegraphics{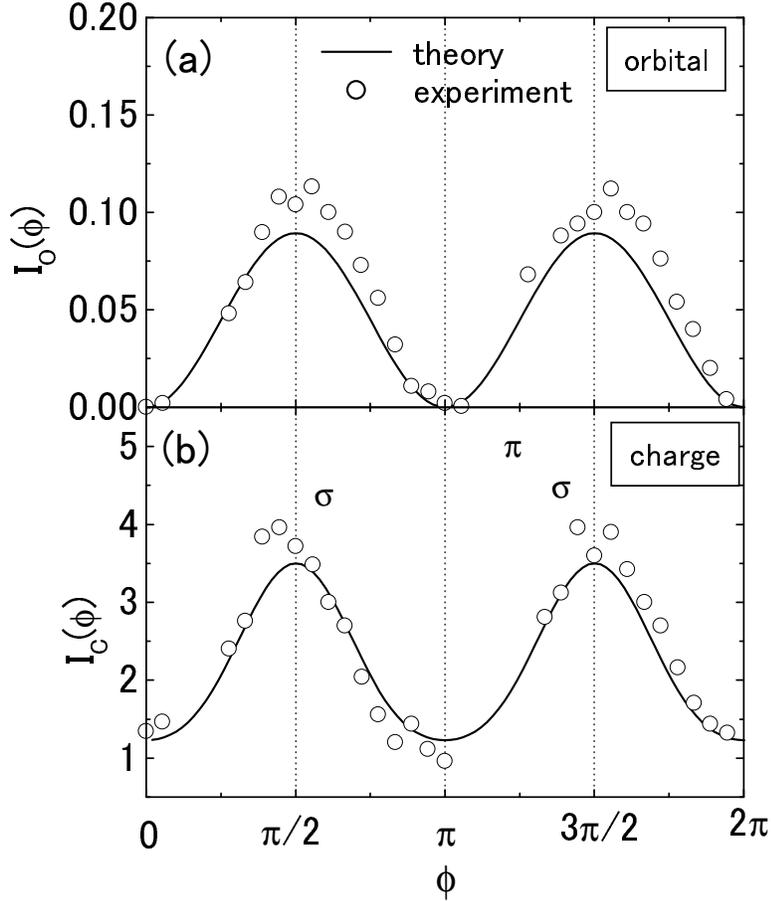}}
\end{center}
        \caption{The azimuthal angle dependence of 
        the scattering intensities at the orbital superlattice reflection (a) and 
        at the charge superlattice one (b) near the edge 
        ($(\omega-(\varepsilon_P-\varepsilon_s))/t_{pd}=-9.6$). 
        The solid curves show the intensities in the  
        $[d_{3x^2-r^2}/d_{3y^2-r^2}]$-type orbital ordered states. 
        The open circles show the experimental data in 
        LaSr$_2$Mn$_2$O$_7$ \cite{wakabayashi}. 
        Absolute values of the experimental data are arbitrary. 
         \cite{ishihara_layer}.
}
\label{fig:azlayer}
\end{figure}
The calculated ASF at the $d_{3x^2-r^2}$ orbital occupied site 
in the $[d_{3x^2-r^2}/d_{3y^2-r^2}]$-type orbital ordered state 
is presented in Fig.~\ref{fig:asflayer}. 
The parameter values are the same as those in Fig.~\ref{fig:asf}, 
except for the hopping integral $t_{\alpha}^{\beta}$ 
between the Mn $4p$ orbitals in Eq.~(\ref{eq:h4p}); 
by considering the symmetry,  $t_{\alpha}^{\beta}$ is parameterized as 
$t_x^x=t_y^y \equiv t_\sigma^\parallel$, $t_z^x=t_z^y=t_x^y=t_y^x \equiv t_\pi^\parallel$ , 
$t_z^z \equiv t_\sigma^\perp$ and $t_x^z=t_y^z \equiv t_\pi^\perp$.
The anisotropy of the hopping integral due to the layered crystal is introduced 
by $t_\sigma^\parallel/t_\sigma^\perp
=t_\pi^\parallel/t_\pi^\perp=4$.
$(\omega-(\varepsilon_P-\varepsilon_s))/t_{pd}=-10$ corresponds to the $K$-edge. 
The weight near the $K$ edge is dominated by Im$\Delta f_{l xx}$ and Im$\Delta f_{l yy}$,  
in contrast to the results for the cubic manganites (Fig.~\ref{fig:asf}) 
where weights of Im$\Delta f_{l yy}$ and Im$\Delta f_{l zz}$ are larger than 
that of Im$\Delta f_{l zz}$. 
This is because the $4p_{x(y)}$ band is broader than the $4p_z$ band 
in the layered crystal structure. 
The $\varphi$ dependences of the scattering intensities at the orbital 
and charge superlattice reflections
were calculated by Eqs.~(\ref{eq:sin2}) and (\ref{eq:coo}), respectively. 
The results together with the experimental data in LaSr$_{2}$Mn$_2$O$_7$ 
are presented in Fig.~\ref{fig:azlayer}. 
The polarization dependences of 
the scattering intensity at the orbital and charge superlattice reflection are 
ascribed to the anisotropies between 
$\Delta f_{xx}$ and $\Delta f_{yy}$ due to the local Coulomb interactions 
and between $\Delta f_{xx}+\Delta f_{yy}$ and 
$\Delta f_{zz}$ due to the effects of the $4p$ band, respectively. 
This figure shows good agreement with theory and experiment. 
In conclusion, from the present comparisons, 
the $[\theta_A/-\theta_A]$-type orbital ordered state is realized in  LaSr$_{2}$Mn$_2$O$_7$,  
and the difference between $\Delta f_{l xx(yy)}$ and $\Delta f_{l zz}$ is dominated 
by the anisotropy of the $4p$ band.  

\subsection{General theoretical framework of RXS - orbital ordering, fluctuation and excitations - }
\label{sec:group}

In the previous two subsections, we introduced the mechanism of the anisotropy of ASF 
in orbital ordered state and 
calculated the energy, polarization and orbital dependence of ASF 
from the microscopic view points. 
The model Hamiltonian includes a number of degrees of freedom 
and several interactions among them. 
This is a standard procedure for analyses of solid state spectroscopy. 
However, the main purpose of RXS is to detect  
the orbital degree of freedom of a $3d$ electron and to reveal their roles 
in correlated electron systems.  
For this purpose, 
we introduce, in this section, a general theoretical framework of RXS 
which is directly connected with the orbital degree of freedom of a $3d$ electron \cite{ishihara_group}. 
The obtained form is applied to study the orbital ordering, fluctuation and excitations. 

Let us start with the differential scattering cross section given in Eq.~(\ref{eq:sigma}). 
This equation is rewritten by using the correlation function 
of the electronic polarizability $\alpha_{\beta \alpha}$:  
\begin{eqnarray}
{d^2 \sigma \over d \Omega d \omega_f}&=&\sigma_T {\omega_f \over \omega_i}
\sum_{\alpha \beta \alpha' \beta'}
P_{\beta' \alpha' } P_{ \beta \alpha}
\Pi_{\beta' \alpha' \beta \alpha}(\omega, \vec K) , 
\label{eq:sigma2}
\end{eqnarray}
where
\begin{eqnarray}
\Pi_{\beta' \alpha' \beta \alpha}(\omega, \vec K)={1 \over 2 \pi}
\int dt e^{i \omega t} \sum_{ll'} e^{-i\vec K \cdot (\vec r_{l'}-\vec r_{l})}
\langle i |\alpha_{l' \beta' \alpha'}(t)^\dagger 
\alpha_{l \beta \alpha}(0)| i  \rangle , 
\label{eq:pialal}
\end{eqnarray}
with 
$\vec K=\vec k_i-\vec k_f$, $\omega=\omega_i-\omega_f$ and 
$P_{\beta \alpha }=(\vec e_{ k_f \lambda_f})_{\beta}   (\vec e_{k_i \lambda_i})_{\alpha}$. 
$\alpha_{l \beta \alpha}(t)$
is the polarizability operator $\alpha_{l \beta \alpha}^{(\omega_i)}$ 
at site $l$ \cite{kolpakov,dmitrienko1,moriya,sakai,shastry};  
\begin{equation}
\alpha_{l \beta \alpha }(t)=e^{i {\cal H}_et} \alpha_{l \beta \alpha}^{(\omega_i) } e^{-i {\cal H}_et} , 
\end{equation}
where ${\cal H}_e$ is the electronic part of the Hamiltonian (Eq.~(\ref{eq:ham})) 
and $\alpha_{l \beta \alpha}^{(\omega_i) }$ is given by 
\begin{equation}
\alpha_{l \beta \alpha }^{(\omega_i)}=i \int_{-\infty}^{0} dt  e^{-i \omega_i t} 
 [ j_{l \beta}(0),  j_{l \alpha }(t) ]  . 
\label{eq:jj}
\end{equation}
This definition of $\alpha_{l \beta \alpha }^{(\omega_i)}$ corresponds to the form 
of $S_{2 \alpha \beta}$ given in Eq.~(\ref{eq:s2}), 
when $\vec j(\vec k)$ is replaced by $\vec j_l$. 

Now the polarizability operator $\alpha_{l \beta \alpha}$ defined at site $l$ 
is expanded in terms of 
the electronic operators of $3d$ electrons  
by utilizing the group theoretical analyses.  
Consider the $O_h$ point symmetry 
around a Mn ion and the doubly degenerate orbitals with the $E_g$ symmetry in each Mn site. 
The pseudo-spin operators are introduced as   
\begin{equation}
T_{l \mu}={1 \over 2} \sum_{\gamma \gamma' \sigma}
d_{l \gamma \sigma}^\dagger (\sigma_\mu)_{\gamma \gamma'}
d_{l \gamma' \sigma} , 
\end{equation}
for $\mu=(0,x,y,z)$ where 
$\sigma_0$ is a unit matrix and $\sigma_{\nu}$ $(\nu=x,y,z)$ are the Pauli matrices. 
$T_{l \mu}$'s represent the charge ($\mu=0$) and orbital ($\mu=x,y,z$) 
degrees of freedom of a $3d$ electron at site $l$ 
and have the A$_{1g}$, E$_{gv}$, A$_{2g}$ and E$_{gu}$ symmetries 
for $\mu=0$, $x$, $y$ and $z$, respectively. 
When $\alpha_{l \beta \alpha}$ associated with a pseudo-spin 
operator at site $l$ is considered, 
$\alpha_{l \beta \alpha}$ is expressed by products of 
$T_{l \mu}$ and a tensor with respect to the polarizations of x ray. 
This tensor has $T_{1u} \times T_{1u}$ symmetry 
being reduced to $A_{1g}+E_{g}+T_{1g}+T_{2g}$. 
Since  the polarizability should have the $A_{1g}$ symmetry, 
$\alpha_{l \beta \alpha}$ is expressed as 
\begin{eqnarray}
\alpha_{l \beta \alpha }
= \delta_{\alpha \beta} I_{A_{1g}} T_{l0} 
+ \delta_{\alpha \beta} I_{E_g} \biggl ( \cos{2 \pi n_\alpha \over 3} T_{lz}
                  -\sin{2 \pi n_\alpha  \over 3} T_{lx} \biggr ) .  
\label{eq:alpht}
\end{eqnarray}
$(n_x,n_y,n_z)=(1,2,3)$ 
and $I_{A_{1g}(E_g)}$ is a constant which is not determined by 
the group theoretical analyses. 
Note the following characteristics in Eq.~(\ref{eq:alpht}): 
(A)  Higher order terms with respect to $T_{l \mu}$ at site $l$, e.g. $T_{lx}T_{lz}$,   
are reduced to the linear terms of $T_{l \mu}$ by using the SU(2) algebra of $\vec T_{l \mu}$. 
(B) The terms including $T_{m \ne l \mu}$, such as $T_{lx}T_{mz}$, are neglected. 
These terms are caused by the higher order processes 
of the electron hopping and the inter-site Coulomb interactions.
(C) $T_{ly}$ describing the magnetic 
octupole moment \cite{khomskii_cm,takahashi_cm,maezono_cm} 
does not appear in Eq.~(\ref{eq:alpht}).  
This is because the coupling constant of $\alpha_{l \beta \alpha}$ 
does not include the $A_{2g}$ symmetry. 
(D) Spin operators are not included. 
This is because the spin-orbit coupling is quenched in 
the $e_g$ orbital states in a Mn ion. 
The magnetic diffraction intensity is expected to be small in 
RXS at $K$-edge in antiferromagnets with linearly polarized x ray. 
(E) Microscopic origins of the $\vec T_{l \mu}$ dependent 
polarizability are introduced in Sec.~\ref{sec:mechanism}, 
i.e. the Coulomb and Jahn-Teller mechanisms, where 
the $1s \rightarrow 4p$ transition  
reflects the $3d$ orbital states through the Coulomb interactions and 
the lattice distortions, respectively. 
However, it is noted that Eq.~(\ref{eq:alpht}) is derived by only considering 
the symmetry of crystal structure and orbitals. 

By using Eq.~(\ref{eq:alpht}),  
Eq.~(\ref{eq:pialal}) is rewritten as   
\begin{eqnarray}
\Pi_{ \beta' \alpha' \beta \alpha}(\omega, \vec K  )&=& \delta_{\beta' \alpha'} \delta_{\beta \alpha}
{1 \over 2\pi} \int dt e^{i\omega t}
\sum_{l l'}e^{-i \vec K \cdot (\vec r_{l'}-\vec r_{l})} \nonumber \\
& \times &
\sum_{\gamma, \gamma'=0,x,z}
I_{\gamma' \alpha'} I_{\gamma \alpha}  
\langle T_{l' \gamma'}(t) T_{l \gamma}(0)  \rangle , 
\label{eq:pitt}
\end{eqnarray}
with 
$I_{0 \alpha}=I_{A_{1g}}$ and 
$I_{x(z) \alpha}=I_{E_g} \cos(-\sin){2 \pi n_\alpha \over 3}$. 
The scattering 
cross section of RXS is directly connected to the dynamical 
correlation function    
of the pseudo-spin operators.
The equation (\ref{eq:sigma2}) with Eq.~(\ref{eq:pitt}) is analogous to the 
scattering cross section in the 
magnetic neutron and conventional x-ray/electron scatterings which 
are represented by the correlation functions of spin and 
charge, respectively. 
In the following. the cross section 
given in Eq.~(\ref{eq:sigma2}) with Eqs.~(\ref{eq:pialal}) and (\ref{eq:pitt}) is applied 
to the scatterings from orbital ordering, fluctuation and excitations. 

(1) {\it Static scattering from orbital ordering}: 
In the static scattering where the scattered x-ray energy is integrated out, 
the cross section is given by the equal-time correlation function: 
\begin{equation}
{d \sigma \over d \Omega} = \sigma_T \sum_{\alpha \alpha'}
P_{\alpha' \alpha'}
P_{\alpha \alpha}
\sum_{\gamma , \gamma'=0,x,z} I_{\gamma' \alpha'}
I_{\gamma \alpha} S_{\gamma' \gamma}(\vec K) , 
\label{eq:static}
\end{equation}
with 
\begin{equation}
S_{\gamma' \gamma}(\vec K)= \sum_{l l'} 
e^{-i \vec K \cdot( \vec r_{l'}-\vec r_{l})}
\langle T_{l' \gamma'} T_{l \gamma}\rangle . 
\label{eq:correlation}
\end{equation}
This form is applicable to 
the scattering from the orbital ordering below the ordering temperature $T_{OO}$. 
Consider the $C$-type AF orbital order of the $[\theta_A/\theta_B]$-type 
which is observed in Pr$_{0.5}$Sr$_{0.5}$MnO$_3$. 
The rotating frame is introduced in the pseudo-spin space at each Mn site: 
\begin{equation}
 {\widetilde T_{l z}}=
\cos \theta_l T_{l z}-\sin \theta_l  T_{l x}  , 
\end{equation}
for $l=A$ and $B$. 
The scattering cross sections at the orbital superlattice reflection 
is obtained from 
Eqs.~(\ref{eq:static}) and (\ref{eq:correlation}) as 
\begin{equation}
{d \sigma  \over d \Omega} \Bigr |_{\sigma \rightarrow \pi}
=\sigma_T N ^2 3 I_{E_g}^2\langle {\widetilde T_{z}} \rangle^2 \sin^2 
\theta_A \sin^2 \varphi \cos^2 \theta , 
\label{eq:opi}
\end{equation}
with $\langle {\widetilde T_z} \rangle ={1 \over 2} \sum_{l=A,B} 
(\langle {\widetilde T_{Az}} \rangle+\langle {\widetilde T_{Bz}} \rangle)$, 
the azimuthal angle $\varphi$ and the scattering angle $\theta$. 
The polarizations of the incident and scattered x ray are chosen to be 
$\sigma$ and $\pi$, respectively, and a relation $\theta_A=-\theta_B$ is assumed. 
This expression corresponds to Eq.~(\ref{eq:sin2}). 
It is worth noting that the cross section 
is proportional to 
$(\langle \widetilde T_z \rangle \sin\theta_A)^2=
\langle (T_{x A}-T_{x B}) \rangle^2$, i.e. 
a square of the AF component of $T_x$. 
Being based on this fact, 
the following two information for the orbital ordering are obtained 
from the experimental data. 
(A) 
Temperature dependence of the orbital order parameter and its critical 
exponent $\beta$ are derived.  
Near $T_{OO}$, the RXS intensity is proportional to 
$(1-T/T_{OO})^{2\beta}$.  
(B) 
The scattering intensity depends on types of the orbital 
ordered state, i.e. a value of $\theta_A$. 
The factor $\sin^2 \theta_A$ becomes maximum in the state of 
$[ {1 \over \sqrt{2}}(d_{3z^2-r^2}-d_{x^2-y^2})/
  {1 \over \sqrt{2}}(d_{3z^2-r^2}+d_{x^2-y^2}) ]$-type  
$(\theta_A=\pi/2)$ 
rather than the $[d_{x^2-z^2}/d_{y^2-z^2}]$- ($\theta_A=\pi/3$)
or $[d_{3x^2-r^2}/d_{3y^2-r^2}]$-types ($\theta_A=2\pi/3$).
A complementary information for the orbital ordered state is obtained 
at the charge superlattice reflection 
$(h\ k\ l)=({2l+1\over 2}\ {2l+1 \over 2}\ n)$. 
The cross section for the $\sigma-\pi$ scattering 
is given by  
\begin{equation}
{d \sigma  \over d \Omega} \Bigr |_{\sigma \rightarrow \pi}
=\sigma_T N ^2 {1 \over 12} I_{E_g}^2 \langle {\widetilde T_{z}} \rangle^2 \cos^2 
\theta_A \sin^2 2\varphi \sin^2 \theta , 
\end{equation}
which is proportional to $(\langle \widetilde T_z \rangle \cos\theta_A)^2=
\langle (T_{z A}+T_{z B}) \rangle^2$. 
By combining RXS study at the orbital and charge superlattice reflections, 
types of the orbital ordered states is expected to be determined.  

(2) {\it Diffuse scattering from the orbital fluctuations}: 
The scattering cross section given in Eqs.~(\ref{eq:static}) and (\ref{eq:correlation}) 
is also applicable  
to study the diffuse scattering near $T_{OO}$. 
The static correlation function $S_{\gamma' \gamma}(\vec K)$ around the orbital superlattice reflection 
near $T_{OO}$ contributes to the critical diffuse scattering in RXS. 
It is worth mentioning that the diffuse scattering in RXS is directly connected  
to the correlation function of the orbital fluctuation  
and each element of $S_{\gamma' \gamma}(\vec K)$  
is obtained by changing the x-ray polarization.
This is in contrast to the previous studies of the x-ray/neutron diffuse scatterings near  
the cooperative Jahn-Teller transition \cite{kataoka_80,terauchi,review_diffuse}.   
\begin{figure}[t]
\begin{center}
        \resizebox{0.75\linewidth}{!}{\includegraphics{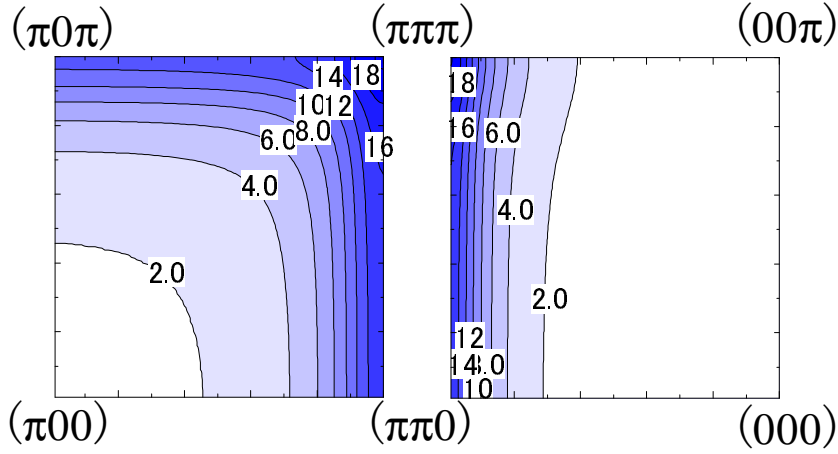}}
\end{center}
        \caption{The intensity contour of the diffuse scattering: 
        $B_{xx}(\vec K)=(NT \chi_0 I_{E_g}^2)^{-1}\sum_{\gamma' \gamma} 
        I_{\gamma' \alpha} I_{\gamma \alpha}S_{\gamma' \gamma}(\vec K)$ 
         at $T=1.05T_{OO}$ \cite{ishihara_group}. }
        \label{fig:diffuse}
\end{figure}
We demonstrate the diffuse scattering in RXS  
calculated from the model where the interaction between orbitals is caused by the 
superexchange-type electronic process \cite{ishihara_eh,kugel72}:   
\begin{eqnarray}
{\cal H}_{J}=&-&2 \sum_{\langle ij \rangle } J_1^l
\biggl ( {3 \over 4} n_i n_j + \vec S_i \cdot \vec S_j   \biggr )
\biggl ( {1 \over 4}  - \tau_i^l \tau_j^l \biggr ) \nonumber \\
     &-&2 \sum_{\langle ij \rangle } J_2^l
\biggl ( {1 \over 4} n_i n_j  - \vec S_i \cdot \vec S_j   \biggr )
\biggl ( {3 \over 4}   + \tau_i^l \tau_j^l +\tau_i^l+\tau_j^l \biggr ) , 
\label{eq:hj}
\end{eqnarray}
where 
\begin{equation}
\tau_i^{l}=\cos \biggl ( {2\pi n_l \over 3} \biggr )T_{iz}-
\sin \biggl( {2 \pi n_l \over 3} \biggr )T_{ix} , 
\end{equation}
with $(n_x, n_y,n_z)=(1,2,3)$. 
$l$ indicates a direction of a bond between site $i$ and 
its NN site $j$. 
$J_{1}^l$ and $J_{(2)}^l$ are the superexchange-type interactions in direction $l$ 
defined by 
$J_1^l=t^{l2}/(U'-I)$ 
and 
$J_2^l=t^{l2}/U$, 
and  
$t^l$ is the hopping integral between the NN Mn sites. 
This model is known to be suitable to describe the spin and orbital states in 
an insulating manganite LaMnO$_3$ \cite{ishihara_eh}.
The intensity contour of the diffuse scattering 
\begin{equation}
B_{\alpha \alpha}(\vec K)={1 \over NT \chi_0 I_{E_g}^2}\sum_{\gamma' \gamma} 
I_{\gamma' \alpha} I_{\gamma \alpha}S_{\gamma' \gamma}(\vec K) , 
\end{equation}
was calculated by the random-phase approximation and the numerical results are presented in Fig.~\ref{fig:diffuse}. 
The scattering cross section is represented by $B_{\alpha \alpha}$ as follows, 
\begin{equation}
{d\sigma \over d \Omega}
=ANT\chi_0I_{E_g}^2\sum_{\alpha' \alpha}P_{\alpha' \alpha'}P_{\alpha \alpha}
B_{\alpha' \alpha}(\vec K). 
\end{equation}
The temperature is chosen to be $T=1.05T_{OO}$.  
A weak anisotropy in the interactions is introduced 
as $J_1^{x(y)}/J_1^z=J_1^{x(y)}/J_1^z=\delta_{x(y)}$
with $\delta_x=1.05$ and $\delta_y=1.025$ 
to remove a degeneracy of the orbital ordered state. 
Below $T_{OO}$, the orbital order occurs at $(\pi \pi \pi)$. 
Strong intensity appears 
along the $(\pi \pi \pi)-(\pi \pi 0)$ direction and other two equivalent ones. 
This characteristic feature is attributed to 
the interaction between orbitals adopted above; 
$J_{xx}(\vec r_l-\vec r_{l'}=a \hat z)=0$. 
This arises from the fact that 
the hopping integral between the $d_{x^2-y^2}$ 
and $d_{x^2-y^2(3z^2-r^2)}$ orbitals is zero in the $z$ direction. 

As introduced in Sec.~\ref{sec:exp}, 
the temperature dependence of the diffuse scattering 
was measured in Pr$_{1-x}$Ca$_{x}$MnO$_3$ around $x=0.4$ \cite{zimmermann_1,zimmermann_2}, 
although the measurement was carried out along one direction in the Brillouine zone. 
By applying the Orstein-Zernike theory to $S_{\gamma' \gamma}(\vec K)$ in Eq.~(\ref{eq:correlation}), 
the peak width around the orbital superlattice reflection is 
given by $\xi(T)^{-1}=\xi^{-1}_0+A(T/T_{OO}-1)^\nu$ 
with constants $\xi_0$ and $A$, and exponent $\nu$. 
The experimental data are well fitted by this equation, although 
more detail measurements   
are required to determine the precise value of $\nu$, as well as the interaction between orbitals. 

(3) {\it Inelastic scattering from orbital excitations}: 
In comparison with the orbital ordering,  
little is known about the orbital excitations and experimental probes to detect them. 
The optical probes have access directly to the orbital excitations, 
although the observation is limited to the zero momentum transfer. 
Recently, a three-peak structure around 150 meV in Raman spectra were observed in LaMnO$_3$ 
and were interpreted to be the scattering from the collective orbital excitations 
termed orbital wave \cite{ishihara_eh,inoue_ow,saitoh_ow,okamoto_ow}. 
Now we focus on the inelastic spectroscopy of RXS
i.e., the resonant inelastic x-ray scattering (RIXS) 
as a probe to detect the orbital excitations. 
This is a momentum resolved probe to detect the bulk 
electronic structures in solids \cite{rixs_rev,gelmukhanov,kotani}. 
Now this experimental technique rapidly progresses accompanied with 
the recent advances of the third-generation synchrotron light source 
\cite{hill_rixs,isaccs,platzman,abbamonte,hasan,ma,tsutsui,tsutsui2,inami}. 

Let us consider the orbital excitations in orbital ordered 
insulating state, such as LaMnO$_3$, 
although the following discussions are applicable to 
an orbital-ordered metal. 
The order parameter in the orbital ordered state 
with the $[3d_{3x^2-r^2}/3d_{3y^2-r^2}]$-type is represented by 
the pseudo-spin as  
\begin{equation}
\langle \vec T_{k} \rangle=
\biggl( {\sqrt{3} \over 4}\delta_{\vec k=(\pi \pi 0)}, 0, 
-{1 \over 4} \delta_{\vec k=(0 0 0)} \biggr), 
\end{equation}
where $\langle \vec T_k \rangle$ is the Fourier transform of 
$\langle \vec T_i \rangle$. 
The orbital excitations 
are represented by deviations of $\vec T_i$ from the above value.
There exist two kinds of the orbital excitations:  
the collective orbital excitation termed the orbital wave and the individual excitation. 
The orbital wave is analogous to the spin wave in the magnetically ordered states 
\cite{cyrot,korovin,ishihara_eh,khaliullin,feiner,brink,perebeinos}. 
When an orbital excitation arises at a Mn site, 
this excitation propagates by 
interactions between orbitals at different sites denoted by $J$. 
Thus, the orbital wave excitations show dispersions and 
their excitation energies are characterized by $J$. 
The energy, dispersion relation and symmetry of the 
orbital wave were calculated 
from the model Hamiltonian such as Eq.~(\ref{eq:hj}) \cite{ishihara_eh,saitoh_ow,ishihara_rixs}. 
The observed peaks by the recent Raman scattering experiment 
in LaMnO$_3$ \cite{saitoh_ow} 
were interpreted to be 
the scattering from the orbital wave at the point $\Gamma$. 
In contrast to the collective excitations, 
the individual orbital excitations show continuum spectra. 
These are analogous to the Stoner excitations in the magnetically ordered states  
and are arise from the electronic excitations from 
the lower Hubbard band (the major orbital band) to 
the upper Hubbard bands (the minor orbital band) across the Mott gap or the charge-transfer gap. 
Thus, the characteristic energy of the orbital excitations is of the order of 
the on-site Coulomb interaction $U'$  
or the charge transfer energy $\Delta$. 

\begin{figure}[t]
\begin{center}
       \resizebox{0.7\linewidth}{!}{\includegraphics{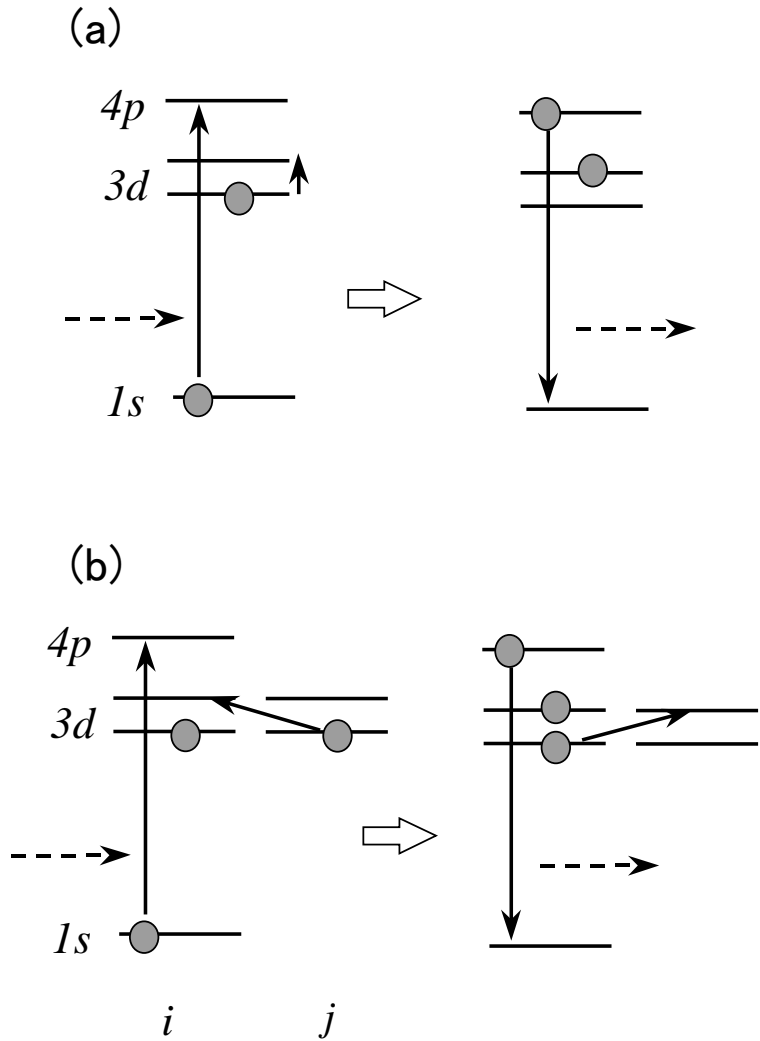}}
\end{center}
        \caption{The scattering processes of the orbital excitations in RIXS. 
        The broken arrows indicate incident and scattered x ray \cite{ishihara_rixs}.}
\label{fig:rixsp}
\end{figure}
Here we introduce RIXS to detect the orbital excitations. 
The resonant effect not only enhances the scattering intensity dramatically 
but also causes several electronic excitations in the intermediate scattering states. 
It was proposed that the following are the possible scattering 
processes from the orbital excitations in RIXS \cite{ishihara_rixs}:  
(A) The incident x ray excites 
an electron from Mn $1s$ orbital to Mn $4p$ one. 
In the intermediate state, the $3d$ electron is excited from the occupied orbital to the unoccupied one  
through the Coulomb interaction between $3d$ and $4p$ electrons. 
Finally, the $4p$ electron fills the core hole by emitting x ray. 
This process is denoted by 
\begin{eqnarray}
| 3d_{\gamma}^1 \rangle +\hbar \omega_i  \rightarrow
| 3d_{\gamma}^1 4p^1  \underline{1s}   \rangle 
\rightarrow 
| 3d_{-\gamma}^1  4p^1 \underline{1s}   \rangle 
\rightarrow
| 3d_{ {- \gamma}}^1   \rangle +\hbar \omega_f , 
\label{eq:kprocess}
\end {eqnarray}
and schematically shown in Fig.~\ref{fig:rixsp}(a).   
The one orbital excitation occurs 
at the site where the x ray is absorbed. 
(B) 
In the intermediate scattering state, 
one hole is created in the Mn $1s$ orbital. 
To screen the core hole potential, 
an electron comes from the NN O $2p$ orbital to the Mn site. 
Due to the hybridization between the O $2p$ and Mn $3d$ orbitals, 
this state strongly mixes with the state 
where the $e_g$ orbitals in 
one of the NN Mn sites ($j$ site) are empty.  
When the $4p$ electron  fills the $1s$ orbital by emitting x ray, 
one of the $3d$ electrons in site $i$ comes back to site $j$. 
This process is denoted by 
\begin{eqnarray}
| 3d_{i \gamma_i}^1 3d_{j \gamma_j}^1\rangle +\hbar \omega_i  \rightarrow 
| 3d_{i  \gamma_i}^1 3d_{i \gamma_i'}^1   
\underline{1s_i} 4p_i^1   \rangle
\rightarrow
| 3d_{i \gamma_i''}^1  3d_{j \gamma_j'}^1  \rangle +\hbar \omega_f ,  
\label{eq:lprocess1}
\end {eqnarray}
and is shown in Fig.~\ref{fig:rixsp}(b). 
Both one- and two-orbital excitations occur in 
$i$ and $j$ sites in this process. 
Since the process (B) arises from the higher order electronic processes 
of the electron hopping and Coulomb interaction,  
the process (A) is expected to be a dominant process of the orbital excitations in RIXS. 

\begin{figure}[t]
\begin{center}
        \resizebox{0.7\linewidth}{!}{\includegraphics{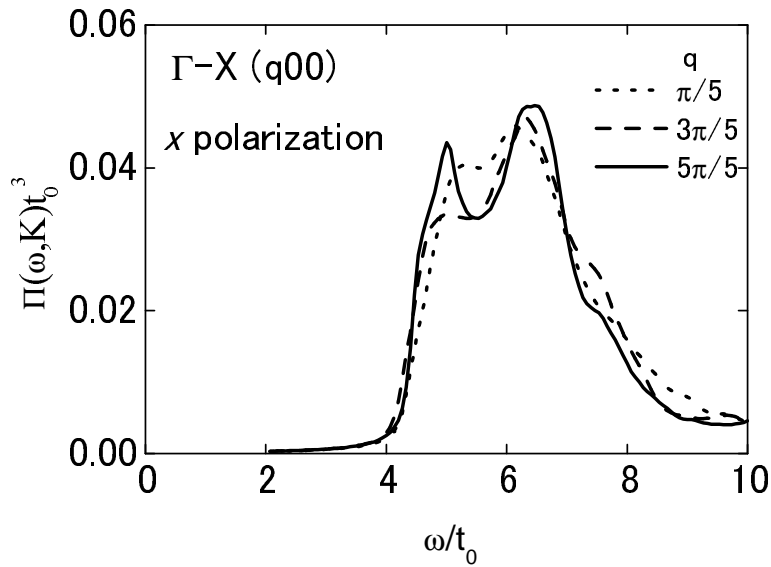}}
\end{center}
        \caption{RIXS spectra $\Pi_{\beta' \alpha' \beta \alpha}(\omega, \vec K)$ for 
        the $[d_{3x^2-r^2}/d_{3y^2-r^2}]$-type orbital ordered state.
        The momentum transfer is $\vec K=(q 0 0)$ ($\Gamma-X$ direction) 
        in the cubic Brillouin zone. 
        The polarization of x ray is chosen to be $\alpha=\alpha'=\beta=\beta'=x$ (the $x$ polarization). 
        $t_0$ is estimated to be about 0.5$\sim$ 0.7eV \cite{kondo}. 
}\label{fig:rixs}
\end{figure}
\begin{figure}[t]
\begin{center}
        \resizebox{0.65\linewidth}{!}{\includegraphics{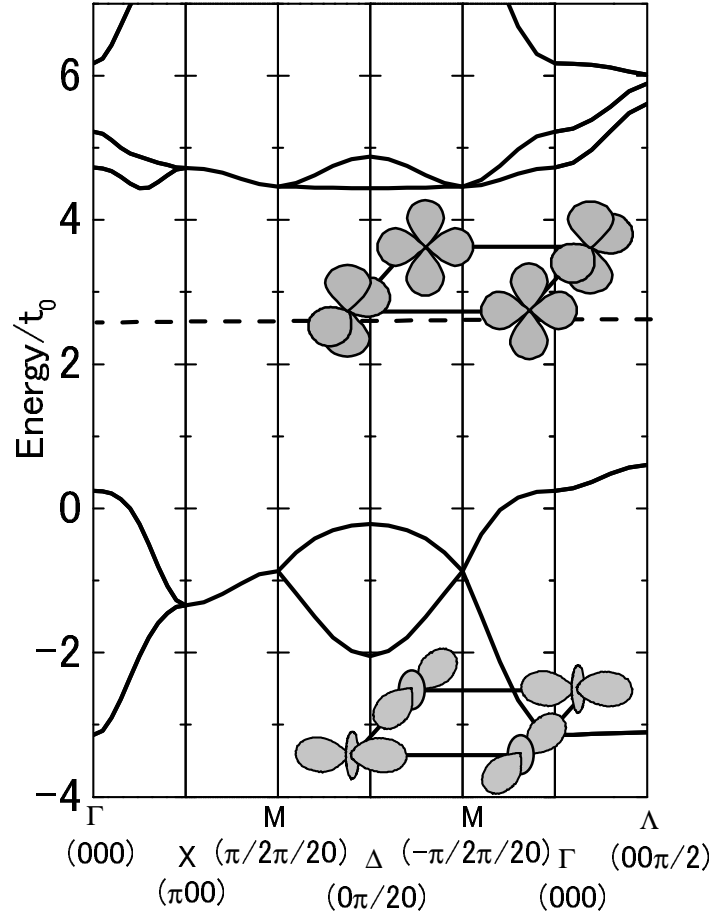}}
\end{center}
        \caption{The electron energy bands for the $[d_{3x^2-r^2}/d_{3y^2-r^2}]$-type orbital ordered state. 
        The dotted line represents the chemical potential located at the center of the occupied and lowest 
        unoccupied bands. The origin of the vertical axis is arbitrary. 
        The inset shows schematic pictures of the orbital ordered states \cite{kondo}. 
 }\label{fig:band}
\end{figure}
The scattering cross section of RIXS from the orbital excitations 
is obtained from Eq.~(\ref{eq:sigma2}) with Eq.~(\ref{eq:pialal}) \cite{kondo}. 
Here, the polarizability operator $\alpha_{l \alpha \beta}$ is expanded by operators 
for the electronic excitations in microscopic viewpoint.
It is considered that the process (A) is dominant and 
the orbital excitations are caused by the off-diagonal Coulomb interaction 
$V_{\gamma -\gamma \alpha}$ given in Eq.~(\ref{eq:w3d4p}).  
By using the Liouville operator method, 
the explicit form of the polarizability is represented as  
\begin{equation}
\alpha_{l \beta \alpha}= 
-{|A_{1s4p}|^2 \over m} \delta_{\alpha \beta} \sum_{\sigma_1 \sigma_2} 
J_{l \alpha \sigma_2}^\dagger J_{l \alpha \sigma_1}
\sum_{\sigma \gamma} 
{T_{l \gamma x \sigma}} D_{ \gamma \sigma \alpha} , 
\label{eq:alpfin}
\end{equation}
with 
$J_{l \alpha \sigma}=P_{l \alpha \sigma}^\dagger s_{l \sigma}$. 
$T_{l  x \sigma}(=T_{l + \sigma}+T_{l - \sigma})/2$ with
$T_{l \mu \sigma}={1 \over 2} \sum_{\gamma \gamma'}
d_{l \gamma \sigma}^\dagger (\sigma_\mu)_{\gamma \gamma'}
d_{l \gamma' \sigma}$
represents a flipping of the pseudo-spin, and 
the factor $D_{\gamma \sigma \alpha}$ is an  
amplitude of the orbital excitations 
from the $3d_\gamma$ orbital to the $3d_{-\gamma}$ one 
by x ray with the polarization $\alpha$. 
The final form of the scattering cross section is 
given by Eq.~(\ref{eq:sigma2}) with the Fourier transform of 
\begin{eqnarray}
\Pi_{\beta' \alpha' \beta \alpha}(t,\vec r_{l'}-\vec r_{l})&=&
{|A_{1s4p}|^4 \over m^2} \delta_{\beta' \alpha'}\delta_{\beta \alpha}
\sum_{\sigma \sigma' \gamma \gamma'}
D_{\gamma' \sigma' \alpha'}^\ast D_{\gamma \sigma \alpha} 
\nonumber \\
&\times&
\langle T_{l'  x \sigma'}(t)^\dagger T_{l x \sigma}(0) \rangle . 
\label{eq:corrr}
\end{eqnarray}
Consider RIXS in LaMnO$_3$ from the individual orbital excitations. 
The correlation function of $T_{l x  \sigma}$ in Eq.~(\ref{eq:corrr}) was calculated 
by applying the Hartree-Fock approximation to the generalized Hubbard model 
with orbital degeneracy. 

The momentum dependence of the RIXS spectra is presented in Fig.~\ref{fig:rixs} \cite{kondo}. 
The energy parameters are the same with 
those in Fig.~\ref{fig:asf}. 
The RIXS spectra do not have an intensity up to about 4$t_0$ 
which corresponds to the Mott gap. 
A weak momentum dependence is seen in the RIXS spectra. 
The energy and momentum dependences of the RIXS spectra reflect 
the electronic band structure presented in Fig.~\ref{fig:band}.
The occupied bands have the $d_{3x^2-r^2}$ and $d_{3y^2-r^2}$ orbital characters, 
the lowest unoccupied bands have the $d_{y^2-z^2}$ and $d_{z^2-x^2}$ orbital ones, 
and the Mott gap opens between these two.  
The main RIXS spectra in Fig.~\ref{fig:rixs} are attributed to the transitions from 
the lower Hubbard band to the upper Hubbard band. 
Note that the lowest unoccupied bands show an almost flat dispersion, 
because electron hopping between the $d_{y^2-z^2}$ and $d_{z^2-x^2}$ orbitals is forbidden in the $ab$ plane. 
This is the reason of the weak momentum dependence of RIXS spectra 
in comparison with that in cuprates \cite{hasan}. 
The RIXS experiment was recently performed in LaMnO$_3$ in SPring-8 \cite{inami}. 
The mainly three peaks were observed around 2.5eV, 8eV and 11eV. 
The lowest peak around 2.5eV shows the momentum and azimuthal angle dependences 
and both are consistent with the theoretical calculations \cite{inami2}. 
It was interpret that this peak structure arises from the individual orbital excitations 
introduced above. 

\section{Summary}
\label{sec:summary}

In this article, we review the recent theoretical and experimental studies of RXS 
in perovskite manganites and related compounds. 
The orbital degree of freedom in magnetic materials 
has been studied since more than 40 years ago. 
The recent discovery of CMR stimulates the intensive investigations based on the standpoint that 
orbital is the third degree of freedom of an electron in addition to spin and charge.  
However, the orbital degree of freedom and the orbital ordering remained hidden. 
The successful observations of the orbital ordering by RXS 
have uncovered this hidden degree of freedom and shown that this is not just theoretical hypotheses. 

Through this review of the theoretical framework of RXS, we show that 
several characteristics of RXS are suitable to detect the orbital ordering. 
In particular, a tensor character of ASF is stressed. 
We introduce possible origins of the anisotropic ASF  
in orbital ordered state. 
The azimuthal angle dependence of the scattering intensity caused by this tensor character 
is crucial to identify the scattering from the orbital ordering. 
Beyond the microscopic description of ASF, 
it has been shown that the scattering cross section of RXS is represented by 
the correlation function of the pseudo-spin operators. 
This expression is applicable to study the orbital order-disorder phase transition. 
In addition to the RXS studies applied to the orbital ordering, 
we introduce the recent attempts to detect the dynamics of 
the orbital degree of freedom by RIXS. 
This method is not only the access to new kinds of excitations in solid 
but also provides a rich information about origin of 
exotic electronic phenomena such as CMR. 
Through the recent rapid progresses of the synchrotron radiation source 
and related experimental technique, 
RXS may develop the orbital physics in correlated electron systems, 
as the neutron scattering has done in magnetic materials. 

\section{Acknowledgements}
The authours would like to thank 
Y.~Murakami, Y.~Endoh, T.~Arima, K.~Hirota, D.~Gibbs, J.~P.~Hill, 
M.~Blume, J.~Mizuki, T.~Inami, S.~Okamoto, H.~Kondo and T.~Hatakeyama. 
This work was supported by the Grant in Aid from Ministry of Education, Culture, 
Sports, Science and Technology of Japan, CREST, and Science and Technology Special 
Coordination Fund for Promoting Science and Technology. 
One of authors (S.M.) acknowledges support of the Hunboldt Foundation.

\end{document}